\documentclass[%
 aip,
 amsmath,amssymb,
 reprint,%
]{revtex4-1}

\usepackage{graphicx}
\usepackage{dcolumn}
\usepackage{bm}

\usepackage[utf8]{inputenc}
\usepackage[T1]{fontenc}
\usepackage{mathptmx}
\usepackage{etoolbox}
\usepackage[caption=false]{subfig}

\usepackage{chemformula}
\usepackage[T1]{fontenc}
\usepackage{graphicx}
\usepackage{amsmath}
\usepackage{graphicx}
\usepackage{setspace}
\usepackage{float}
\usepackage{makecell}
\usepackage{multirow}
\usepackage{amsmath}
\usepackage{empheq}
\usepackage{longtable}
\usepackage{booktabs}
\usepackage{adjustbox}
\usepackage[percent]{overpic}
\usepackage[colorinlistoftodos]{todonotes}
\usepackage[colorlinks=true, allcolors=blue]{hyperref}

\makeatletter
\def\@email#1#2{%
 \endgroup
 \patchcmd{\titleblock@produce}
  {\frontmatter@RRAPformat}
  {\frontmatter@RRAPformat{\produce@RRAP{*#1\href{mailto:#2}{#2}}}\frontmatter@RRAPformat}
  {}{}
}%
\makeatother
\begin{document}

\preprint{AIP/123-QED}

\title{A Combined Tight Binding with Machine Learning Potential Model for Magnesium Compounds}
\author{Jiwen Yu}
\affiliation{Department of Materials, Imperial College London.}
\author{Arash A. Mostofi}%
\affiliation{Department of Materials, Imperial College London.}

\author{Andrew Horsfield}
\email{a.mostofi@imperial.ac.uk; a.horsfield@imperial.ac.uk}
\affiliation{Department of Materials, Imperial College London.}

\date{\today}

\begin{abstract}
We present a model for magnesium-based systems that combines Density Functional Tight Binding (DFTB) with MACE, a machine learning interatomic potential (DFTB+MACE). In this model, the conventional repulsive potential pair potential, is replaced by a many-body MACE potential. The MACE component of the model is trained on the difference between Density Functional Theory (DFT) energies and forces and the corresponding DFTB values, but neglecting the pair potential contribution. Using this model we performed structural relaxation of \ch{MgO}-\ch{CO2} adsorption systems, molecular dynamics calculations of water clusters and phonon spectrum calculations of stable fcc-\ch{MgO} and metastable bcc-\ch{MgO} structures. We compare the performance of our model with a pure MACE model and with DFT. We demonstrate that the DFTB+MACE model achieves improved accuracy relative to DFTB with a pair potential, in many cases with only a moderate increase in computational cost. In addition, it can provide electronic structures that most of the machine learning potentials cannot. The training dataset, originally developed for MACE, may not fully represent all regions of the potential surface we may encounter during simulations. Expanding the dataset for a wider potential surface is expected to further enhance predictive accuracy of DFTB+MACE model. Overall, the resulting DFTB+MACE framework enables simulations at length and time scales beyond the reach of first-principles methods while retaining an explicit description of electronic structures, making it particularly attractive for understanding and predicting processes in materials that require an accurate description of charge transfer.
\end{abstract}

\maketitle


\section{Introduction}
Density Functional Tight Binding (DFTB)\cite{porezag1995construction}\cite{seifert1996calculations} is an approximate method derived from Density Functional Theory (DFT)\cite{kohn1965self}. It can provide approximate electronic structures as it is a quantum mechanical method, while calculations are significantly more computationally efficient than DFT, making it suitable for static and dynamical simulations at large length-scales and time-scales.

Magnesium is the lightest engineering metal, and its common corrosion product \ch{MgO} and \ch{Mg(OH)2} can also be used for \ch{CO2} storage due to their chemical reactivity. In our previous work\cite{yu2025tight} we developed a DFTB parameter set for magnesium-based compounds and used it to study the adsorption of \ch{CO2} and \ch{H2O} on \ch{MgO} and \ch{Mg(OH)2} surfaces. Whilst our DFTB model was able to provide a satisfactory description of structural and electronic properties, it still exhibits limitations, which we elaborate below.

In a conventional DFTB parameterisation \cite{wahiduzzaman2013dftb,oliveira2015dftb,knaup2007initial,van2018tight,bodrog2011automated,hellstrom2013scc,hourahine2025recent}, atomic orbitals are typically generated from isolated single atom calculations with an additional confinement potential. The difference between the DFT reference energies and the DFTB electronic energy is incorporated into a repulsive potential, which usually takes the form of a pairwise function represented by a cubic spline interpolation. However, this approach has several limitations: $($i$)$ oscillations may arise in the fitted function, $($ii$)$ the form of the function is often too rigid, $($iii$)$ overfitting may occur, and $($iv$)$ the limited transferability restricts its applicability to only a subset of structures. Kandy \emph{et al}. developed curvature constrained splines to solve this problem. \cite{ammothum2021curvature} They tried to fit the curvature rather than the value of function itself. They found that a pair potential cannot be ``one-fits-all'' because it cannot simultaneously describe different coordination environments. This work highlighted an important limitation of the accuracy of DFTB and that the repulsive potential is missing many-body terms. Machine learning many body potentials may be a good solution to this problem.

Machine Learning Interatomic Potentials (MLIPs) for constructing accurate potential energy surfaces have developed at a remarkable speed in recent years. At their core, MLIPs typically consist of two parts: a descriptor and a machine learning algorithm.  \cite{pinheiro2021choosing} The descriptor can range from simple atomic position information to a complex basis set to describe the atomic environment \cite{behler2007generalized, bartok2010gaussian,drautz2019atomic}. The machine learning algorithms include non-neural network methods\cite{bartok2010gaussian}, simple feedforward neural networks and more complex neural networks such as E(3)-equivariant graph neural networks (e3nn). \cite{batatia2022mace,csanyi2025design, geiger2022e3nn} Several studies have explored the integration of machine learning methods with DFTB to improve the accuracy of the latter. The repulsive term, or the difference between DFT and DFTB, is either described by MLIPs or is defined as the error and obtained by $\Delta-$learning. For example, Martin \emph{et al}. \cite{stohr2020accurate} introduced SchNET neural networks to DFTB to improve the predicted structure of organic molecules. Goldman \emph{et al}. \cite{goldman2021semi,goldman2023enhancing} introduced a MLIP called ChiMES \cite{lindsey2017chimes} as the many-body repulsive potential and applied it to silicon polymorphs, \ch{TiH2} and organic molecules containing \ch{H}, \ch{C}, \ch{N} and \ch{O}. This method was also applied to organic molecules at the water-platinum interface \cite{wang2025development}. 
Recently, e3nn models have been introduced into DFTB frameworks\cite{sandonas2026advancing} to study large organic molecules. The authors considered three different e3nn methods including SpookyNet\cite{unke2021spookynet}, Allegro\cite{musaelian2023learning} and MACE \cite{batatia2022mace,Batatia2022Design} for large organic molecules: they found that MACE performs the best. However, their research does not cover periodic systems and has not been applied to condensed matter. 

In this work, we develop a DFTB+MACE framework for periodic materials by combining the explicit electronic description of self-consistent DFTB with the flexibility of equivariant machine learning potentials. Note that our aim here is not to produce the most accurate model possible (though that remains a long term objective) but to demonstrate a more limited objective. We wish to compare, for a given available training set, the relative performance of DFTB, MACE and DFTB+MACE. Using \ch{MgO} and \ch{Mg(OH)2} based systems as representative examples, we demonstrate that the proposed approach substantially improves the accuracy of DFTB while retaining its ability to describe charge redistribution. The resulting model achieves near-DFT accuracy for energies and forces at a computational cost suitable for large-scale atomistic simulations. These results highlight the potential of DFTB+MACE as an efficient route towards modeling complex materials processes that require both long spatiotemporal scales and an explicit treatment of electronic effects.

\section{Theory and Computational Method}
\subsection{DFTB}
The total energy of DFTB can be derived from a Taylor expansion of the Hohenberg-Kohn\cite{hohenberg1964inhomogeneous} total energy around a chosen reference density $n^{(0)}(\vec{r})$. The difference between the actual density $n(\vec{r})$ and the reference density is the density fluctuation term $q(\vec{r})$, where $n(\vec{r}) = n^{(0)}(\vec{r}) + q(\vec{r})$. The reference density is normally the sum of single atomic charge density and is isotropic. Therefore, the DFTB energy expanded to the second order term around the reference density is:

\begin{align}
    \label{Edftb}
    E^{\text{DFTB}}[n^{(0)}(\vec{r})+ q(\vec r)] =&\underbrace{E[n^{(0)}]}_{E^{(0)}} \notag +\underbrace{\int\frac{\delta E}{\delta n(\vec{r})} \bigg|_{n^{(0)}} q(\vec{r})d\vec{r}}_{E^{(1)}} \notag\\
    &+\underbrace{\frac{1}{2}\iint\frac{\delta^2 E}{\delta n(\vec{r}) \delta n(\vec{r}')} \bigg|_{n^{(0)}} q(\vec{r})q(\vec{r}')d\vec{r}d\vec{r}'}_{E^{(2)}} \notag \\
    &+... \notag \\
    =&E^{(0)}+E^{(1)}+E^{(2)}+...
\end{align}

To define expressions for $E^{(1)}$ and $E^{(2)}$ we introduce a basis set in terms of which we can expand molecular orbitals. The theory is based on the ansatz of a linear combination of atomic orbitals $\varphi_{\alpha i}$ (LCAO) for the molecular orbitals $\psi_n$
\[
\psi_n = \sum_{\alpha i} c_{\alpha i, n}\varphi_{\alpha i}
\]
where $c_{\alpha i,n}$ is an expansion coefficient of molecular orbital $n$ for orbital $i$ of atom $\alpha$. Note that this takes us to the Kohn-Sham formulation of DFT\cite{kohn1965self}. The formulas for the first three terms then are:
\begin{empheq}[left=\empheqlbrace]{align}
    E^{(0)} &= \sum_{\alpha} E^{\text{atom}}_\alpha + \frac{1}{2} \sum_{\alpha\neq \beta} V_{\alpha\beta}(R_{\alpha\beta}) + E^{\text{MB}}_{\text{XC}}[n^{(0)}]
    \label{E0} \\[6pt]
    E^{(1)} &= \underbrace{\sum_{n}f_{n} \epsilon_n^{(0)}}_{\mathrm{band\,energy}}
    - \underbrace{\sum_{\alpha i} f_{\alpha i} H^{(0)}_{\alpha i, \alpha i}}_{\mathrm{atomic\,term}}
    \label{E1} \\[6pt]
    E^{(2)} &= \frac{1}{2} \sum_{\alpha lm}\sum_{\beta l'm'} Q_{\alpha lm}Q_{\beta l'm'} B_{\alpha \beta l m l' m'}(\vec{R}_{\alpha \beta})
    \label{E2}
\end{empheq}

$E^{(0)}$ contains isolated atomic contributions $\sum_\alpha E^{atom}_\alpha$, electrostatic attraction and repulsion between atoms $\frac{1}{2} \sum_{\alpha\neq \beta} V_{\alpha\beta}(R_{\alpha\beta})$ and a many-body exchange-correlation term $E^{\text{MB}}_{\text{XC}}$. The combination $E^{(1)}+E^{(2)}$ we call the electronic energy. $E^{(1)}$ contains the band energy and an atomic term.  $f_n$ and $f_{\alpha i}$ are the molecular occupation number and atomic orbital occupation number respectively. $\epsilon_n^{(0)}$ is the energy for molecular orbital $n$, and it depends on the onsite energy, hopping integral and overlap integral. $H^{(0)}_{\alpha i,\alpha i}$ is the onsite energy for orbital $i$ of atom $\alpha$. $E^{(2)}$ corresponds to electrostatic interactions between perturbed atoms. The quantity $Q_{lm}$ is a charge density moment and $B_{\alpha \beta lml'm'}$ is the Madelung structure constant, which is a function of atomic separation $\vec{R}_{\alpha \beta}$. Detailed derivations can be found in our previous work. \cite{yu2025tight, boleininger2016gaussian}

Normally, the many-body contribution from exchange and correlation in $E^{(0)}$ is neglected, letting $E^{(0)}$ be short range and pairwise, and so is often referred to as the pair potential term. However, this simplification does not consider the chemical environment of a bond. For example, consider compressing the carbon-carbon bond in ethane, and stretching the carbon-carbon bond in ethene, to make them the same length. The $E^{(0)}$ for the carbon-carbon part for these two systems would be the same, which need not be the case in more accurate calculations. For a large system, the errors may accumulate. In this paper, the pair potential approximation for $E^{(0)}$ is relaxed, and we use a machine learned many-body representation (MACE) instead.

Since the total energy is expressed as a sum of independent contributions ($E^{(0)}$,  $E^{(1)}$ and  $E^{(2)}$), the contribution to the total energy, $E^{(0)}$, forces $F^{(0)}$ and stresses $\sigma^{(0)}$ carried by the MACE potential are then given by
\begin{empheq}[left=\empheqlbrace]{align}
    E^{(0)} &= E^{\text{DFT}} - (E^{(1)} + E^{(2)})
    \label{E0a} \\[6pt]
    F^{(0)} &= F^{\text{DFT}} - (F^{(1)} + F^{(2)})
    \label{F0a} \\[6pt]
    \sigma^{(0)} &= \sigma^{\text{DFT}} - (\sigma^{(1)} + \sigma^{(2)})
    \label{S0a}
\end{empheq}
where $E^{(1)}$, $F^{(1)}$, $\sigma^{(1)}$, $E^{(2)}$, $F^{(2)}$ and $\sigma^{(2)}$ are computed using the DFTB model: see equations \ref{E1} and \ref{E2}, while $E^{\text{DFT}}$, $F^{\text{DFT}}$ and $\sigma^{\text{DFT}}$ are calculated by DFT.

\subsection{MACE}
We here present the theory of MACE in some detail as we wish to explain the hyperparameters that we need to adjust to optimize the training process. For a system with $N$ atoms, the total energy of this system can be represented as a sum of individual atomic energies: 
\begin{equation}
    \label{Et}
    E = \sum_\alpha E_{\alpha}
\end{equation}
The energy of atom $\alpha$ depends on its local atomic environment and can be written as:
\begin{equation}
    \label{Ei}
    E_{\alpha}(\vec{\sigma}) = E_{\alpha}(\vec{r}_{1\alpha}, \vec{r}_{2\alpha},...,\vec{r}_{N\alpha})
\end{equation}
where $\vec{\sigma}$ represents a set of displacement vectors $\vec{r}_{\beta\alpha}$. 
MACE combines the ACE descriptor with an E(3)-equivariant message passing neural network (E(3)-MPNN). The use of an E(3)-equivariant neural network (E3NN) enforces E(3)-equivariance with respect to rotations, translations and inversions. Atomic features are represented as irreducible representations (irreps) of the O(3) group and transformed according to their angular momentum order, also known as irreps order. There are three main stages in MACE, which are \textit{embedding}, \textit{interaction} and \textit{product}. See reference \cite{batatia2022mace} for full details.

In the embedding stage, there is an initialization of node features $h_{\alpha k l_1m_1}$, where $k$ is the channel index, $l_1$ is the angular momentum index and $m_1$ is the magnetic quantum number index or the components of the $l^{th}$ irrep. The size of the node features are controlled by the hyperparameters \colorbox{lightgray}{\texttt{num\_channels}} and \colorbox{lightgray}{\texttt{max\_L}}, which are the number of channels for each irrep type and the highest order of irreps in the node features. In the first step, the atomic number is used as the feature for each node, so there is only $h_{\alpha k 00}$ in the node feature at the start. 

There is also an angular embedding to generate spherical harmonic functions $Y_{l_2 m_2}(\hat{r}_{\alpha \beta})$, where $l_2$ and $m_2$ are angular momentum and magnetic quantum numbers, and $\hat{r}_{\alpha \beta}$ is the unit vector connecting atom $\alpha$ to atom $\beta$. The maximum order of the spherical harmonic term is controlled by hyperparameter \colorbox{lightgray}{\texttt{max\_{ell}}}. There is another embedding called radial embedding to form a series of basis sets \{ $j_0(r_{\alpha \beta})$ \}. The influence of neighbours will decay as their separations becomes larger, and the output of the radial embedding can act as the basis of radial functions which will appear in the next stage. The basis set can be constructed from several possible sets of functions, including Bessel functions and Chebyshev polynomials.

In the interaction stage, an intermediate feature $\hat{h}^{(t)}_{\alpha k l_1 m_1}$ is formed from a linear combination of node features. The intermediate feature is coupled with spherical harmonics by an equivariant tensor product $h^{(s)}_{\alpha k l_1 m_1} \otimes Y_{l_2 m_2} $ and forms an output irrep with order $l_3$. The radial functions are generated by a multi-layer perceptron (MLP), and the number of radial functions that we need should be equal to the number of output irreps. 

Now let us select a central atom $\alpha$, with its neighbours being $\beta$. We can then build our one-particle basis:
\begin{align}
    \phi^{(t)}_{\beta k l_3 m_3} = \sum_{l_1 l_2 m_1 m_2} C^{l_3 m_3}_{l_1 m_1 l_2 m_2} R^{(t)}_{ kl_1 l_2 l_3}(r_{\beta\alpha}) \hat{h}^{(t)}_{\beta k l_1 m_1} Y_{l_2 m_2}(\hat{r}_{\beta\alpha})
\end{align}
where $C^{l_3 m_3}_{l_1 m_1 l_2 m_2}$ are Clebsch-Gordan coefficients. Then the atom base for atom $\alpha$ is built as a linear combination of all one-particle basis functions:
\begin{align}
    A^{(t)}_{\alpha k l_3 m_3} = \sum_{k'} \sum_\beta W^{(t,l_3)}_{kk'\alpha  \beta} \phi^{(t)}_{\beta k' l_3 m_3}
\end{align}

In the product stage, the atom bases will be further coupled together through equivariant tensor products. Through setting the hyperparameter \colorbox{lightgray}{\texttt{correlation}}, which controls the maximum body order of the interaction, $A_{l_1 m_1}$ (two-body interaction), $A_{l_1 m_1} \otimes A_{l_2 m_2}$ (three-body interaction) \dots and $A_{l_1 m_1} \otimes A_{l_2 m_2} \otimes ... \otimes A_{l_\nu m_\nu}$ ($\nu +1$-body interaction) can be obtained. After the coupling, the output will be used to generate the cluster basis as a linear combination,
\begin{align}
    B^{(t)}_{\alpha k LM} = \sum _{\{ l_\nu m_\nu \}}C_{\{l_\nu m_\nu \}}^{LM} \sum_{k'} \prod_{\zeta = 1}^{\nu}W^{(t,l_\zeta)}_{kk'} A^{(t)}_{\alpha k' l_\zeta m_\zeta}
\end{align}
where $B^{(s)}_{\alpha k LM}$ is an element of cluster basis, $\zeta$ is an index to distinguish $l$ in different atom basis functions. The coefficient $C_{\{l_\nu m_\nu \}}^{LM}$ is a Clebsch-Gordan coefficient. It should be also pointed that the tensor product with irreps order higher than the maximum allowed angular momentum for features will be ignored. The "message" is defined as a linear combination of cluster basis functions, and the features can be updated as a linear combination of the message and previous features:
\begin{align}
    m^{(t)}_{\alpha kLM} &= \sum_{k'} W^{(t, L)}_{kk'}B^{(t)}_{\alpha k' LM}\\
    h^{(t+1)}_{\alpha k L M} &= \sum_{k'} W^{(t, h)}_{k k'} h^{(t)}_{\alpha k' LM} + \sum_{k'} W^{(t,m)}_{kk'}m^{(t)}_{\alpha k' LM}
\end{align}
The values $L$ and $M$ will become $l_1$ and $m_1$ again in the next layer. In the traditional E3NN method, higher-order many-body correlations are built progressively through multiple interaction layers. In contrast, the descriptor in MACE enables the model to capture many-body interactions in the first layer. The number of MPNN layers is controlled by the hyperparameter \colorbox{lightgray}{\texttt{num\_interactions}}. Because the energy is rotationally invariant, in the last layer the energy can be obtained by
\begin{align}
    E = \text{MLP}(h^{(t)}_{\alpha k00})
\end{align}
To ensure consistency, the forces are generated by direct differentiation of the energy rather than being generated from a multilayer perceptron. 

In summary, the five hyperparameters mentioned in the theory are summarized in table \ref{table1}. The symbols that represent those hyperparameters are used below.

\begin{table*}[]
    \caption{\label{table1} The symbols, meaning and their name in the MACE code for 5 different hyperparameters studied in this work.}
    \begin{tabular}{c c c c c}
        Symbol & Meaning & Name in MACE version 0.3.13\\
        \hline
        $k$ & Number of channels      & \colorbox{lightgray}{\texttt{num\_channels}} \\
        $T$ & Number of MPNN layers & \colorbox{lightgray}{\texttt{num\_interactions}} \\
        $\nu$ &Up to ($\nu+1$)-body interactions included & \colorbox{lightgray}{\texttt{correlation}} \\
        $L_{\text{max}}$ &Highest-order of equivariant features& \colorbox{lightgray}{\texttt{max\_L}} \\
        $l_{\text{max}}$ &Maximum angular momentum of spherical harmonics& \colorbox{lightgray}{\texttt{max\_ell}} \\
    \end{tabular}
\end{table*}
\subsection{Calculation Details}
The PLATO package \cite{soin2011efficient} was used to generate atomic orbitals, the hopping and overlap integral tables and the Hubbard parameters. The Perdew-Burke-Ernzerhof (PBE) functional \cite{perdew1996generalized} was used to describe the electronic exchange and correlation behaviour, and separable dual-space Gaussian pseudo-potentials \cite{goedecker1996separable,hartwigsen1998relativistic,krack2005pseudopotentials} were used to describe the electron-nucleus interactions. The radii of the atomic orbitals ($r_0$) were set to twice an element's covalent radius \cite{porezag1995construction}.

After the DFTB parameters were obtained, they were used to calculate the electronic energy for the machine learning datasets. The datasets include atomic structures that only contain \ch{Mg}, \ch{C}, \ch{O} and \ch{H} from the Materials Project MPtrj dataset \cite{deng2023materials}, which is the dataset used for MACE foundation model\cite{batatia2022mace, Batatia2022Design,batatia2025foundation}, the structures that only contain \ch{H}, \ch{C}, \ch{N}, \ch{O}, \ch{F}, \ch{Mg}, \ch{Al}, \ch{Si}, \ch{P}, \ch{S}, \ch{Cl}, \ch{Cu} and \ch{Br} from the MatPES dataset \cite{kaplan2025foundational}, and some other self-generated xyz structures for \ch{MgO} and \ch{Mg(OH)2}. The self-generated xyz structures include fcc-\ch{MgO} and bcc-\ch{MgO} supercells with random displacements applied to the atoms, 
as well as \ch{CO2} and \ch{H2O} adsorbed on \ch{MgO} and \ch{Mg(OH)2} clusters. There are 15,396 structures in total. The forces and energies were all recalculated with Quantum ESPRESSO \cite{giannozzi2009quantum} using the PBE functional and projector augmented-wave (PAW) pseudopotentials \cite{kresse1999ultrasoft} to ensure consistency across the whole training set. The k-point sampling is fully converged. The dispersion correction \cite{grimme2004accurate} was not considered. 

MACE version 0.3.13 \cite{batatia2022mace,Batatia2022Design} was adopted for MACE model training. Five different hyperparameters were studied as shown in table \ref{table1}. The parameter $k$ was set to values from 8 to 64 with an interval of 8, the parameter $T$ was set to values from 1 to 4, the parameter $\nu$ was set to values from 1 to 5, the parameter $L$ was set to values either 0 or 1 and the parameter $l$ was set to values from 1 to 5. The force-energy weighting ratio was set to 10:1. The same random seed was used to split the dataset into 13861 configurations as a training set and 1535 configurations as a validation set for each combinations of hyperparameters. The role of validation is to evaluate the transferability of this model and prevent the model from overfitting. The dataset was not explicitly divided into a separate test set. Instead, the calculation cases, including bulk, surface, adsorption static and dynamical calculations, were used as a test set to evaluate the model performance. The training was run on a Quardro RTX6000 GPU performed with 32-bit floating point arithmetic.

The generated models were used for static relaxation, molecular dynamics and phonon spectrum calculations. All molecular dynamics simulations were carried out at 300 K within the NVT ensemble and a temperature rescaling thermostat. A finite difference method was used for phonon spectrum calculations \cite{kresse1995ab}, and a supercell with lattice constant larger than 10 Å was created to minimize the interactions between periodic images. A displacement of $\pm~0.01$ Å is applied in three Cartesian directions. The code used for the finite difference method of phonon spectrum calculations can be found through the GitHub link provided in the Data and Code Availability section. In this work, the non-analytical correction (NAC) \cite{pick1970microscopic} was not added. To study bulk water on a brucite surface a $4\times 4\times 5$ supercell of brucite was created to form a slab, and a $12.4\times 10.74\times 15$ \AA$^3$ water layer containing 53 water molecules was placed on the surface of the brucite. The density of the water was set to be equal to the experimental equilibrium density for bulk water. All DFTB-MD calculations were carried out at 300 K within the NVT ensemble and using a temperature rescaling thermostat. The time step was set to be 1 fs. Only the $\Gamma$ point was used for k-point sampling as the system is large enough. All DFTB calculations were performed on Intel Core i5-12500 with 6 cores.

\section{Results and Discussion}
\subsection{Accuracy and complexity analysis}
The MACE model was trained on DFT total energies and forces, while the MACE correction in the DFTB+MACE model was trained on the difference between the DFT total energies and forces and the DFTB electronic energies and forces (see equation \ref{E0a}-\ref{S0a}). To distinguish our MACE model trained with a limited training set from the standard MACE model, we refer to this model as the restricted MACE model (denoted as $\text{MACE}_\text{R}$) throughout the remainder of this paper. We again note that we are looking for the relative performance of our three models for a given training set. In the final evaluation the DFTB+MACE predictions are constructed as the sum of the DFTB electronic term and the MACE correction, and are compared directly with the DFT reference. We evaluate model accuracy using the mean absolute error (MAE). We tested 419 different hyperparameter settings and detailed information for each hyperparameter setting can be found in the Supporting Information. Due to the large amount of data required to explore completely five different variable hyperparameters, we simplify the analysis by considering only three representative categories: simple ($T$=1, $k$=8, $L_{\text{max}}$=0, $\nu$=1, $l_{\text{max}}$=1), moderate ($T$=2, $k$=32, $L_{\text{max}}$=1, $\nu$=2, $l_{\text{max}}$ = 2), and complex ($T$=4, $k$=64, $L_{\text{max}}$=1, $\nu$=3, $l_{\text{max}}$ = 3) models.

\begin{figure}[htbp]
    \centering
    \begin{overpic}[width=0.45\textwidth]{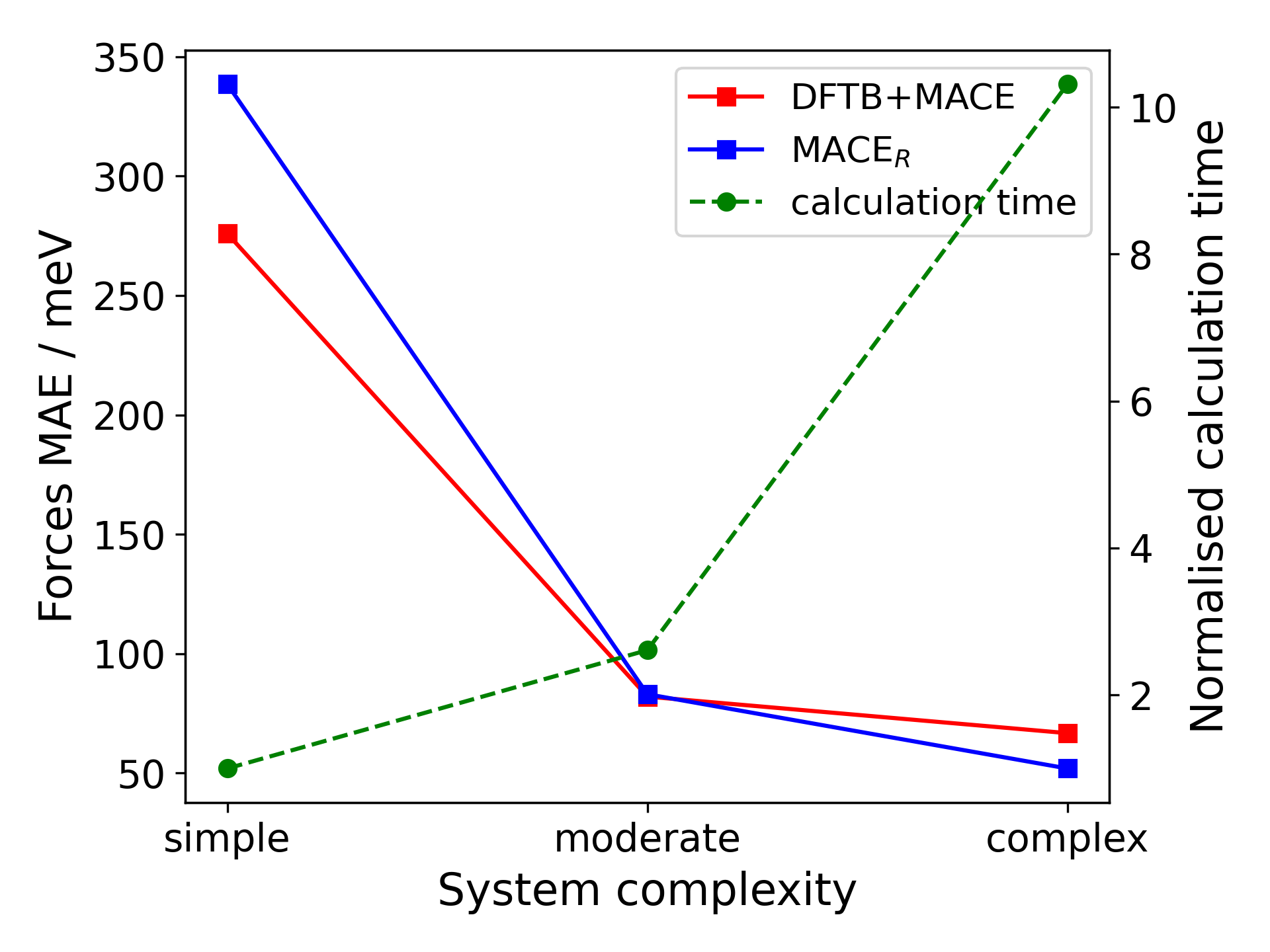}
    \end{overpic}
    \caption{The mean absolute error of forces and normalized calculation time for the \ch{Mg}-\ch{C}-\ch{O}-\ch{H} system against different hyperparameter settings. The time for the simple model is set to be 1, and other values are expressed as multiples of that time.}
    \label{fig:T}
\end{figure}

\begin{figure*}[!htbp]
    \centering
    \subfloat{%
        \begin{overpic}[width=0.32\textwidth]{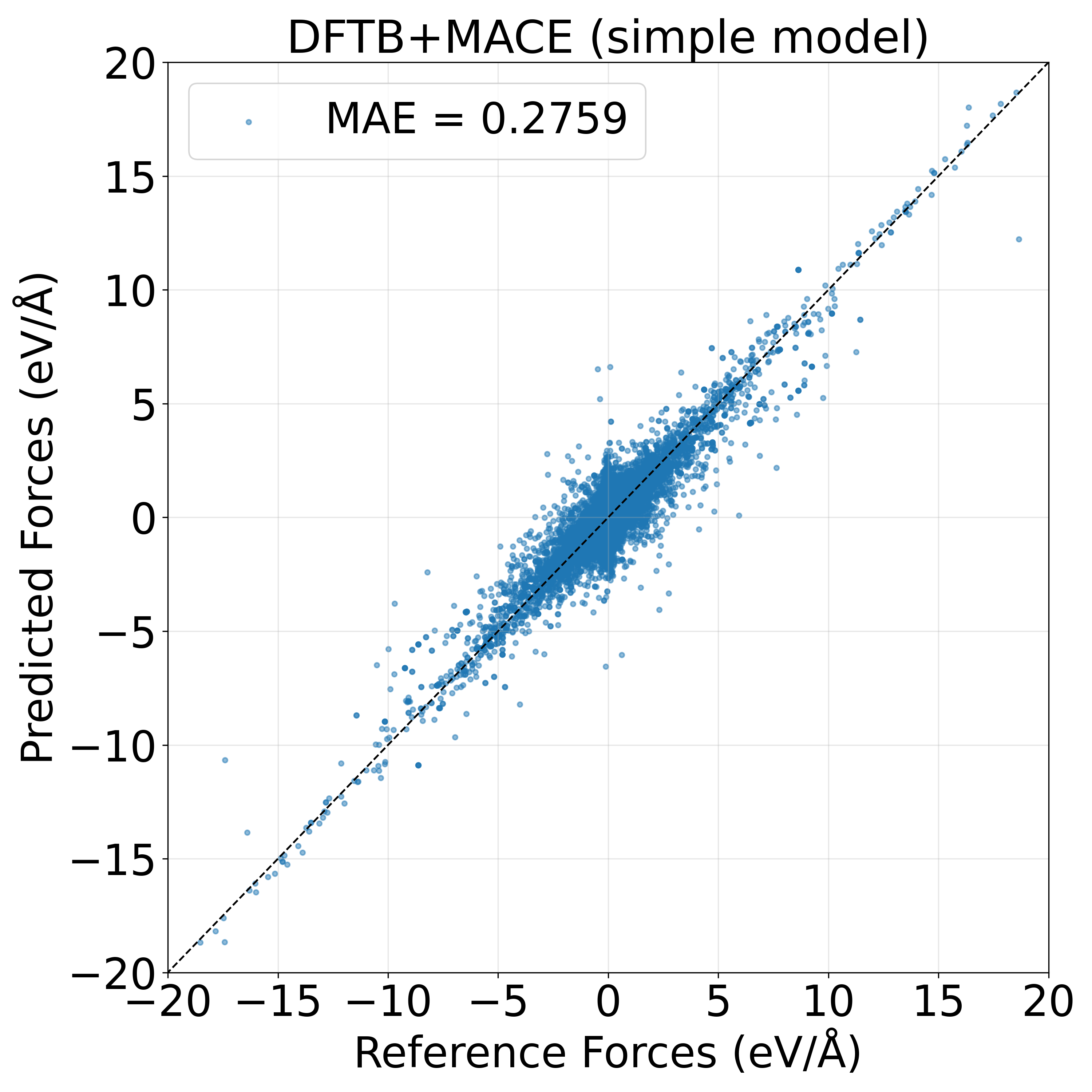}
            \put(0,98){\textbf{(a1)}}
        \end{overpic}%
    }
    \hfill
    \subfloat{%
        \begin{overpic}[width=0.32\textwidth]{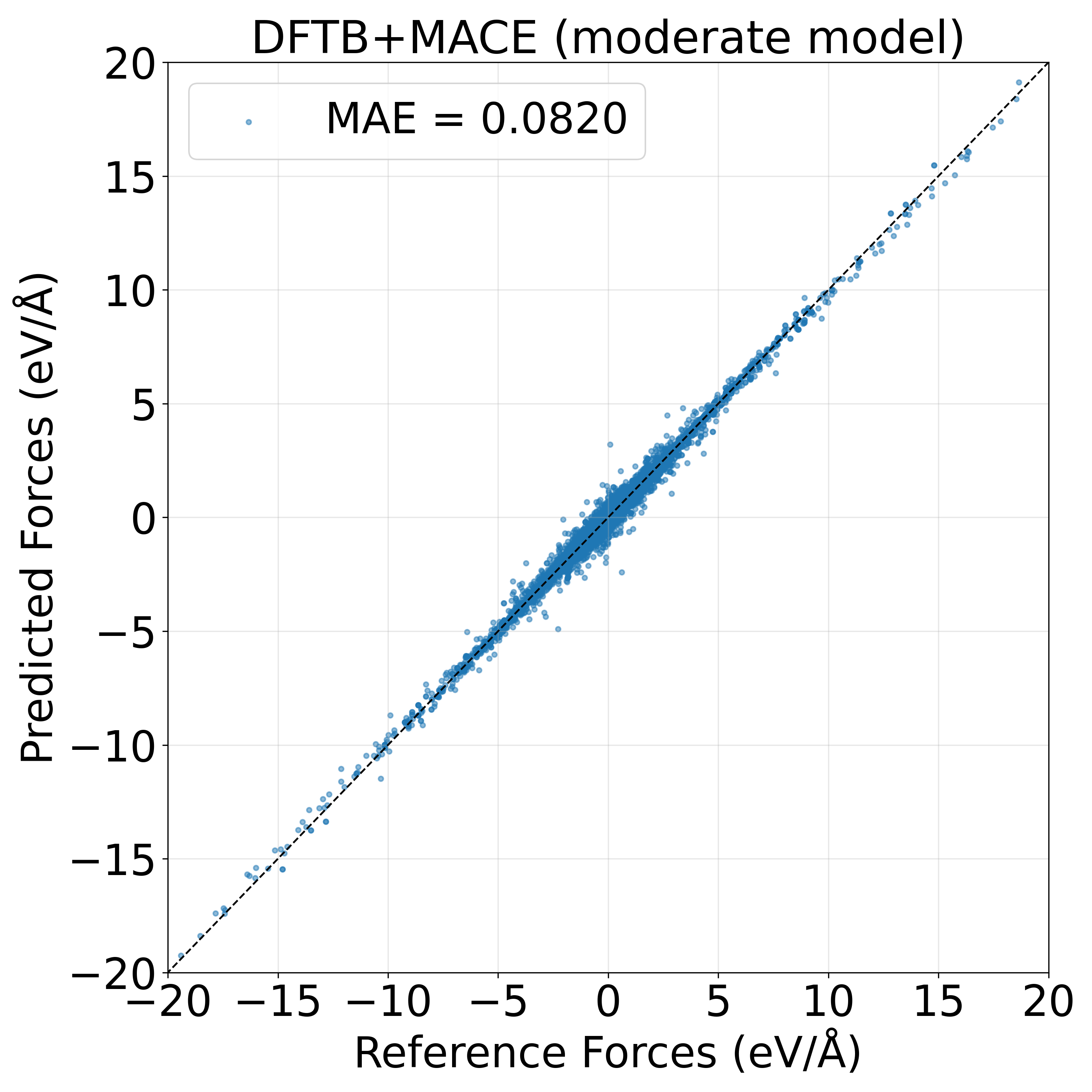}
            \put(0,98){\textbf{(b1)}}
        \end{overpic}%
    }
    \hfill
    \subfloat{%
        \begin{overpic}[width=0.32\textwidth]{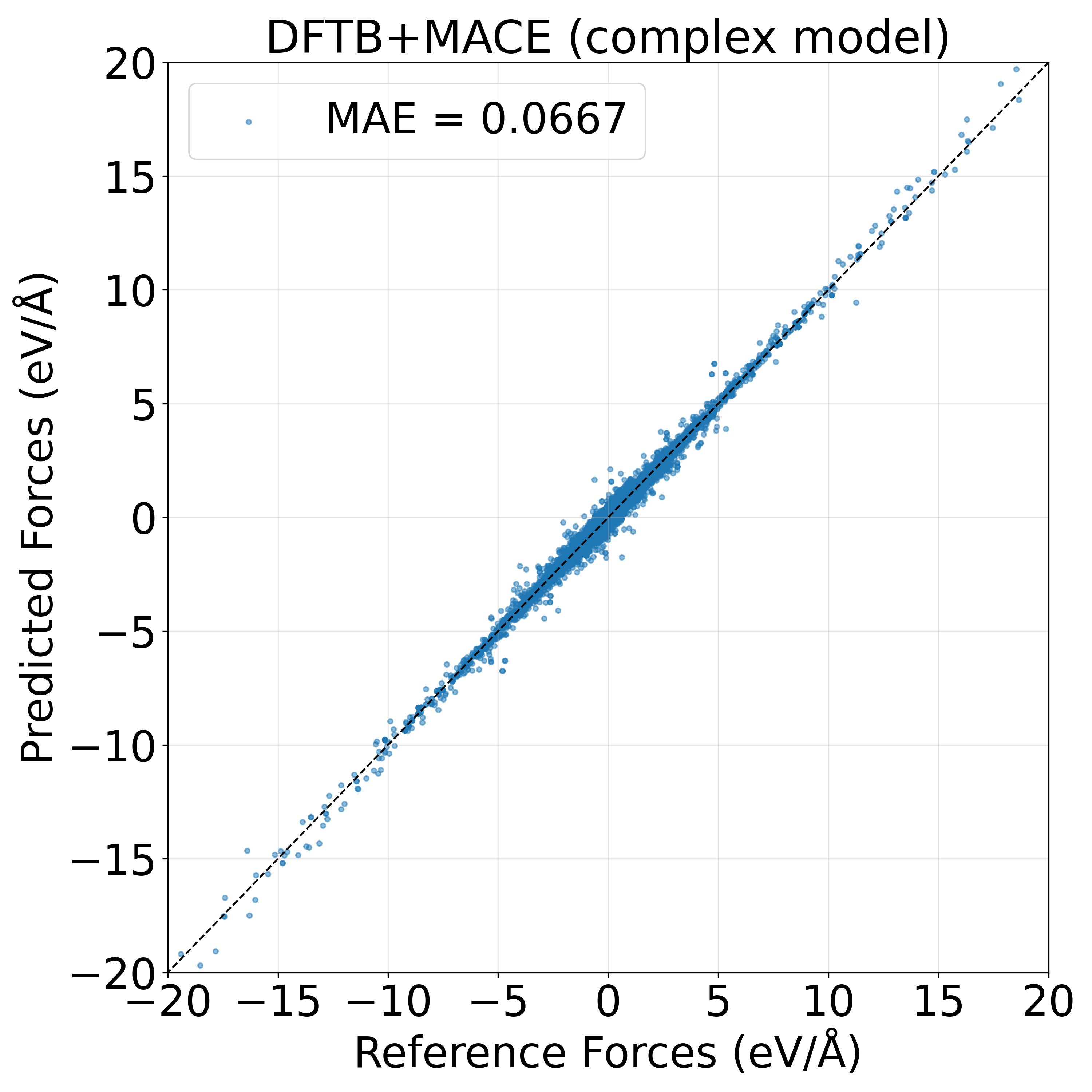}
            \put(0,98){\textbf{(c1)}}
        \end{overpic}%
    }
    \hfill
    \\[1ex]
    \subfloat{%
        \begin{overpic}[width=0.32\textwidth]{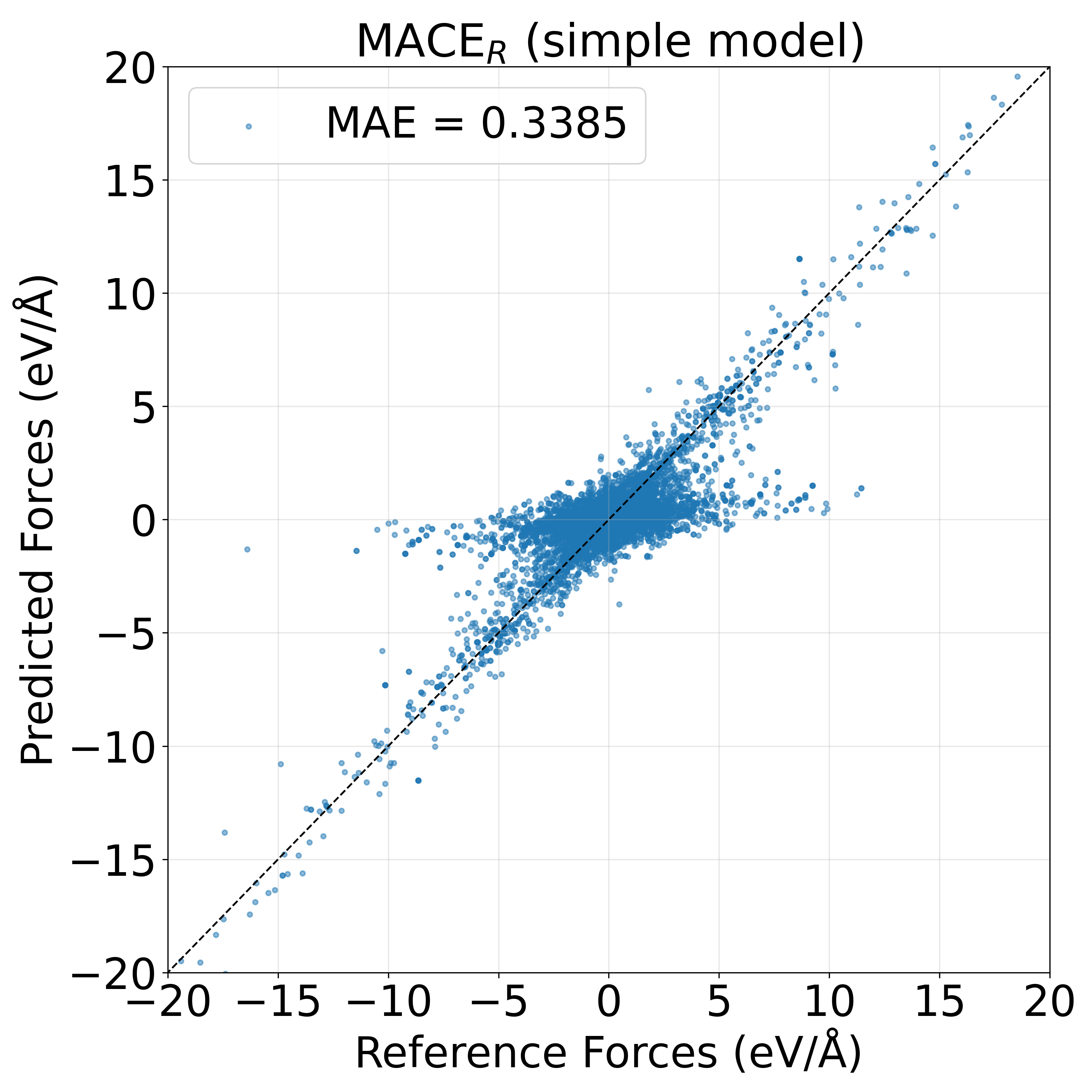}
            \put(0,98){\textbf{(a2)}}
        \end{overpic}%
    }
    \hfill
    \subfloat{%
        \begin{overpic}[width=0.32\textwidth]{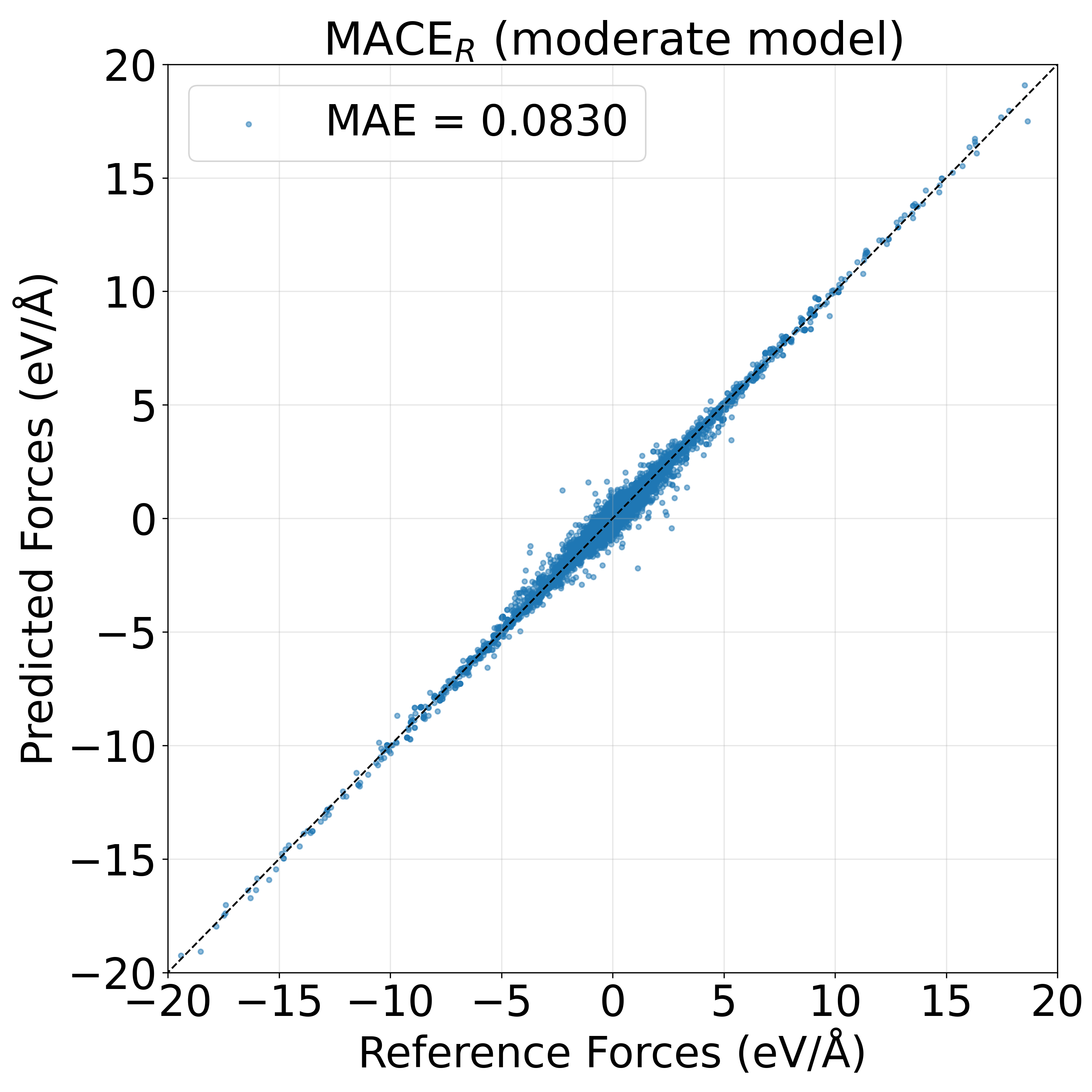}
            \put(0,98){\textbf{(b2)}}
        \end{overpic}%
    }
    \hfill
    \subfloat{%
        \begin{overpic}[width=0.32\textwidth]{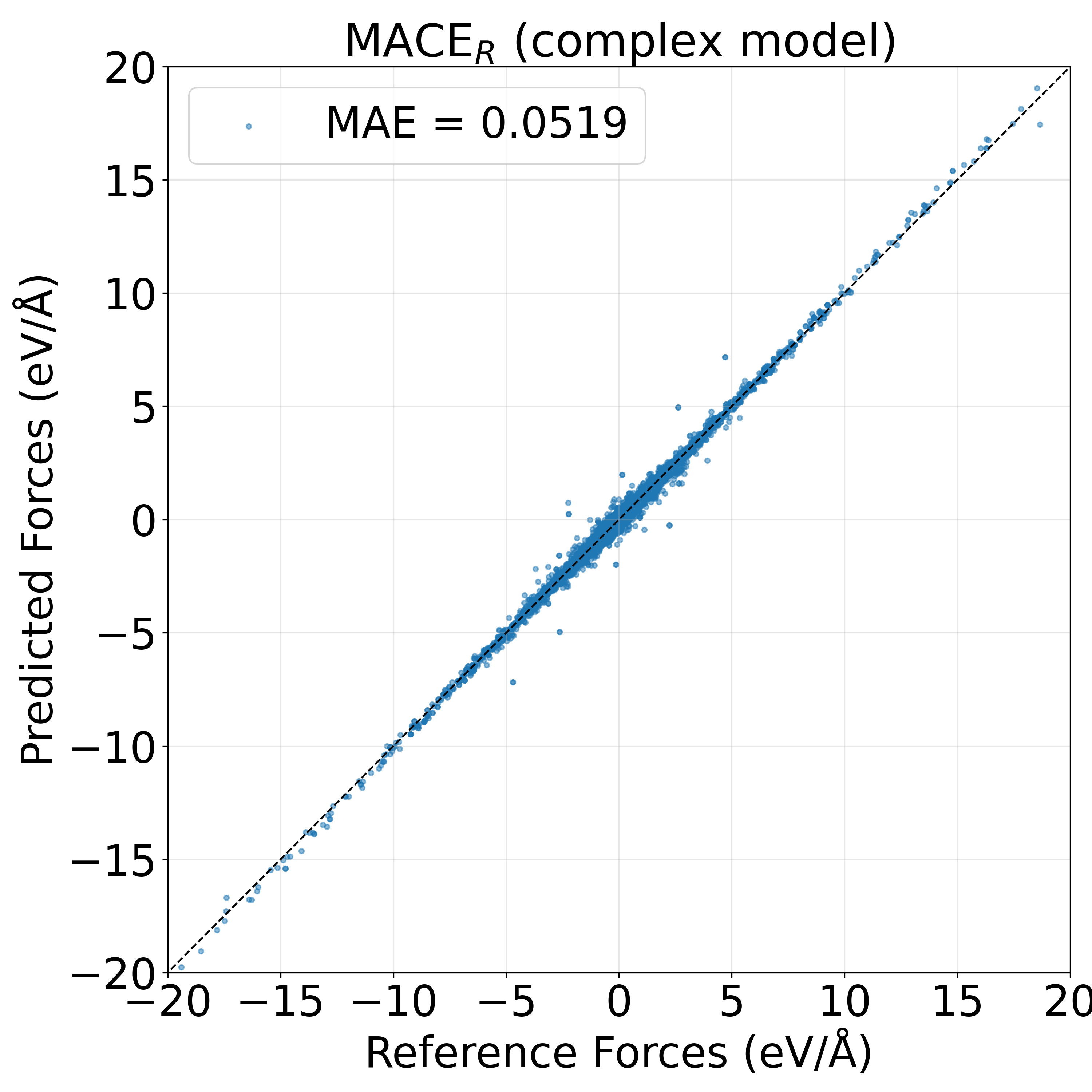}
            \put(0,98){\textbf{(c2)}}
        \end{overpic}%
    }
    \caption{The predicted force values for the \ch{Mg}-\ch{C}-\ch{O}-\ch{H} system against the reference values for the DFTB+MACE model (first row) and $\text{MACE}_\text{R}$ model (second row) with (a) simple models, (b) moderate models and (c) complex models.}
    \label{fig:R}
\end{figure*}

Figure \ref{fig:T} shows the plot of mean absolute error of forces and training time against different hyperparameter settings. Figure \ref{fig:R} shows the comparison between the DFTB+MACE model and the $\text{MACE}_R$ model with different hyperparameter settings. It is found that for a DFTB+MACE model a simple MACE contribution achieves better performance (MAE$_F$ = 275.9 meV/Å) than the corresponding $\text{MACE}_\text{R}$ model (MAE$_F$ = 338.5 meV/Å), and the correlation plot (Figure \ref{fig:R}) also shows the model is more reliable. One reason is that the atomic reference density is isotropic, which implies that $l=0$ is sufficient to describe the repulsive term of the system. Higher order $l$  terms contribute mainly to correct errors arising from the approximations in DFTB, including the exchange and correlation term in $E^{(0)}$, the errors in $E^{(1)}$ and $E^{(2)}$ and the neglected higher order terms. In comparison, the $\text{MACE}_\text{R}$ model cannot capture any information with higher order irreps. That partly explains why the DFTB+MACE model can perform better than a $\text{MACE}_\text{R}$ model with relatively modest hyperparameter settings. For the moderate model and the complex model, the MAE$_F$ for the DFTB+MACE model is 82.0 meV/Å and 66.7 meV/Å respectively, while the values of MAE$_F$ for the corresponding $\text{MACE}_\text{R}$ model are 83.0 meV/Å and 51.9 meV/Å respectively. Thus, for more complex hyperparameter settings, the $\text{MACE}_\text{R}$ model performs comparably well or even slightly better than the DFTB+MACE model. When a more complex set of hyperparameters is used, the 
error in the DFTB electronic term begins to be apparent. In our previous work it was shown that the valence band width in DFTB is smaller than that of the DFT reference, and the gap between the valence band and the lower valence band is larger than the DFT reference due to the limited basis set selected in DFTB. Thus the DFTB+MACE model accuracy might be further improved by introducing crystal field and three-centre integral corrections to generate a higher quality DFTB parameter set. While the code exists to include these terms, optimizing their use is a topic for future work.

While only the simple, moderate, and complex categories are shown here for clarity, the complete analysis covering all five hyperparameters yields the same overall trend. The computational cost increases rapidly and non-linearly with increasing $l$: see figures 5-a1 and 5-a2 in the SI. For many-body interactions, which are related to the hyperparameter $\nu$, both models show a clear improvement in accuracy on going from two-body to three body terms, while further increases in body order lead to only marginal improvements. In contrast, increasing the body order leads to a significant increase in computational cost: see figures 5-b1 and 5-b2 in the SI. The MAE decreases approximately exponentially with increasing number of MPNN layers ($T$) and the number of channels ($k$): see figure 5-c1, 5-c2, 5-d1 and 5-d2 in the SI. In principle, the computational cost is expected to increase linearly with both hyperparameters. However, for small channel sizes, the increase is not clear, which may relate to GPU utilization, bandwidth limitation, package framework and so on. 

\subsection{Static Calculations}

To further assess the accuracy and transferability of our models, the static properties of \ch{MgO}, including cohesive energy as a function of lattice constant, phonon spectrum, surface energy and adsorption problems, were studied by the moderate model. 

\begin{figure*}[htbp]
    \centering
    \subfloat{%
        \begin{overpic}[width=0.45\textwidth]{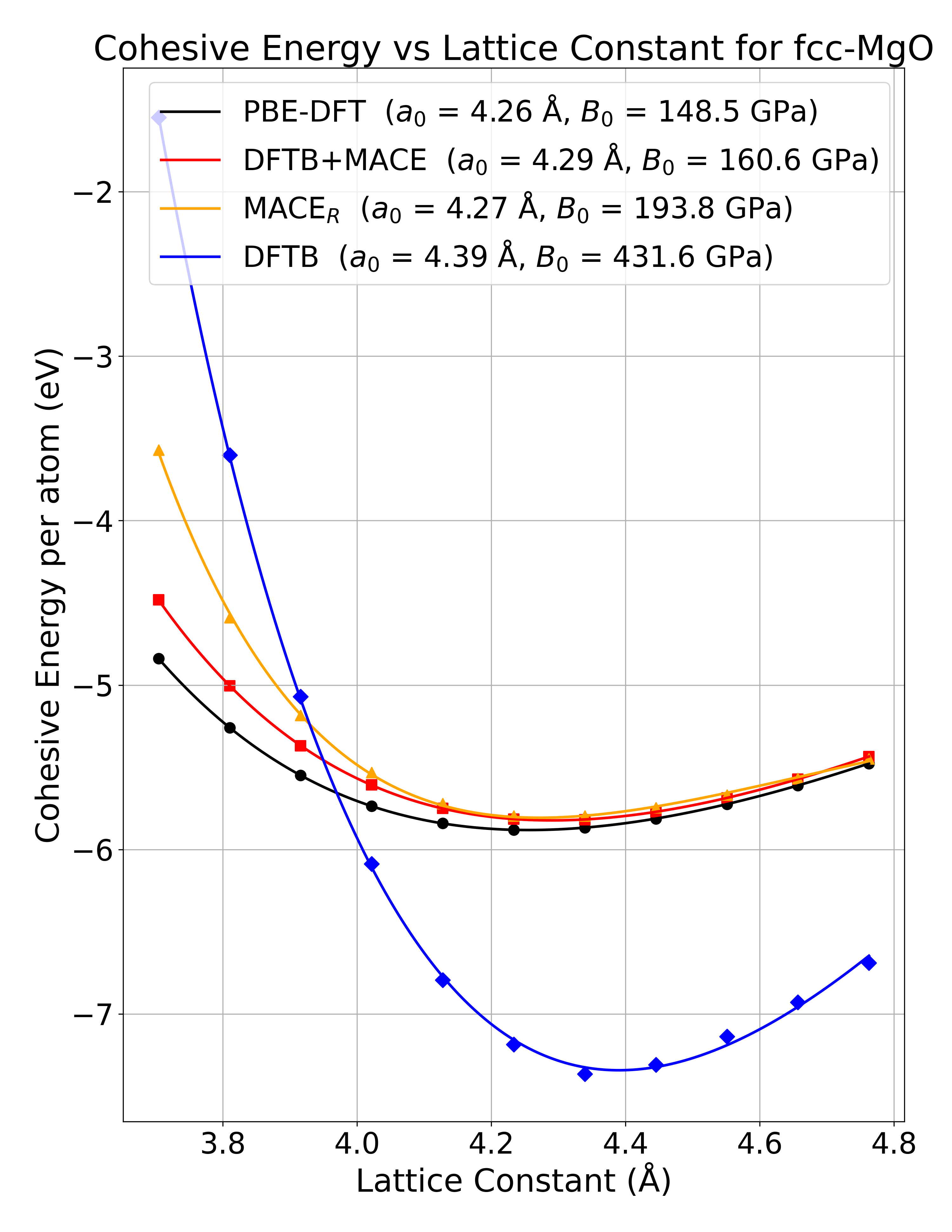}
            \put(2,90){\textbf{(a)}}
        \end{overpic}%
    }
    \hfill
    \subfloat{%
        \begin{overpic}[width=0.45\textwidth]{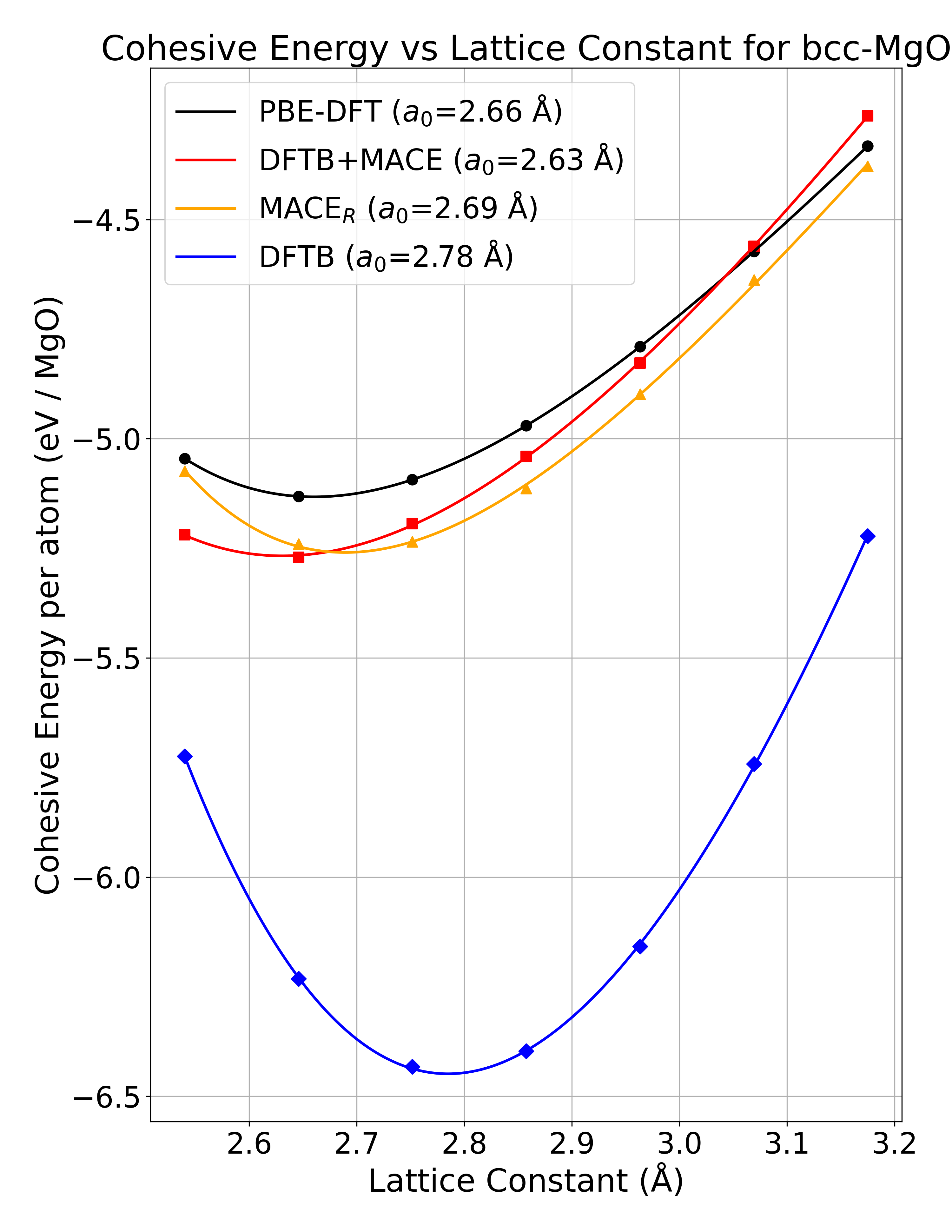}
            \put(2,90){\textbf{(b)}}
        \end{overpic}%
    }
    \caption{The cohesive energy of (a) fcc-\ch{MgO} and (b) bcc-\ch{MgO} against the lattice constant. The moderate models are used for DFTB+MACE model and MACE$_{\text{R}}$ and the curve is fitted by a third-order Birch-Murnaghan Equation of State.}
    \label{fig:Ecurve}
\end{figure*}

Figure \ref{fig:Ecurve} shows the cohesive energy curve of fcc and bcc \ch{MgO} for different models or parameter sets. The lattice constant ($\text{a}_0$) and bulk modulus ($\text{B}_0$) are labelled on the plot. It can be seen that there is a large improvement in both the lattice constant and the shape of the binding energy curve compared with DFTB. However, while the model curves agree well with the DFT data where the training dataset covers the relevant atomic configurations, the behavior outside this data range is  less certain. The details of the atomic configurations in the dataset can be found by following the \textbf{github} link and from figures 1-4 in the \textbf{Support Information}.

For fcc-\ch{MgO}, the overall trend from the DFTB+MACE model is better than for the $\text{MACE}_\text{R}$ model. However, around the equilibrium point they are quite similar. One reason could be that the training data set does not cover configurations with non-equilibrium lattice constants, in which case and model has to make an estimate. In addition, there is a more negative contribution from the electronic energy from tight binding model when the lattice constant becomes smaller, which becomes increasingly difficult for the MACE part to compensate.

The same issues are apparent for bcc-\ch{MgO}. The $\text{MACE}_\text{R}$ model exhibits a similar behaviour: in regions with a small lattice parameter not covered by the training data the energy rises too rapidly. For the DFTB+MACE model, the appearance of the electronic part makes the gradient similar to that of the PBE-DFT reference.

Even after adding configurations generated by random perturbations to the equilibrium lattice, this issue was not fully resolved. Although lattice constants far from the equilibrium value are not particularly important for simulations at relatively low temperatures, they become essential for high-temperature MD simulations. Therefore, the training dataset that includes configurations with a range of lattice constants is necessary for high temperature MD simulations. This is a project for future work.

From the energy-lattice constant curve we can obtain the equilibrium lattice constant and bulk modulus, which characterize the structural stability. However, this only considers isotropic compression and expansion, and now we further analyse the dynamical stability by looking at the phonon spectrum, which is related to the full set of second-order derivatives of the energies with respect to atomic displacement. 

\begin{figure}[htbp]
    \centering
    \begin{overpic}[width=0.45\textwidth]{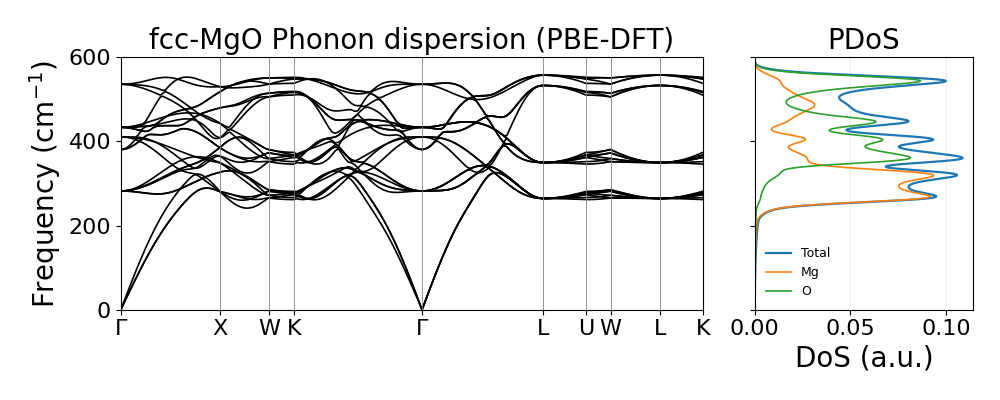}
        \put(1,35){\textbf{(a)}}
    \end{overpic}
    \par\vspace{1ex}
    \begin{overpic}[width=0.45\textwidth]{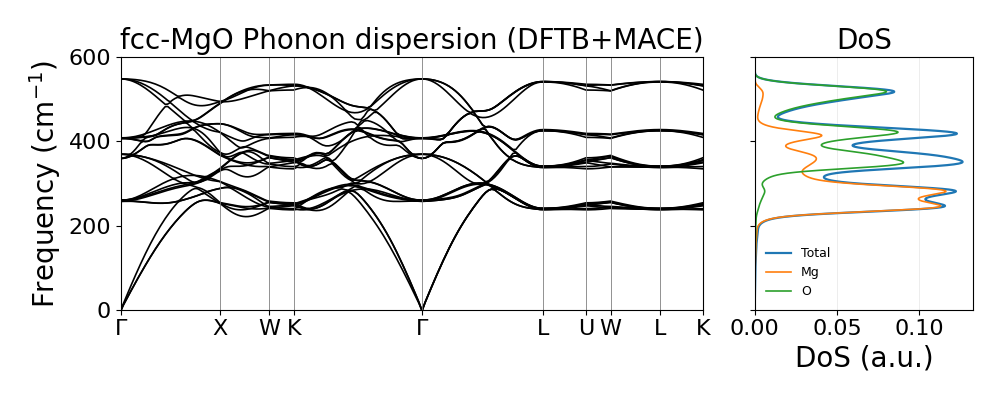}
        \put(1,35){\textbf{(b)}}
    \end{overpic}
    \par\vspace{1ex}
    \begin{overpic}[width=0.45\textwidth]{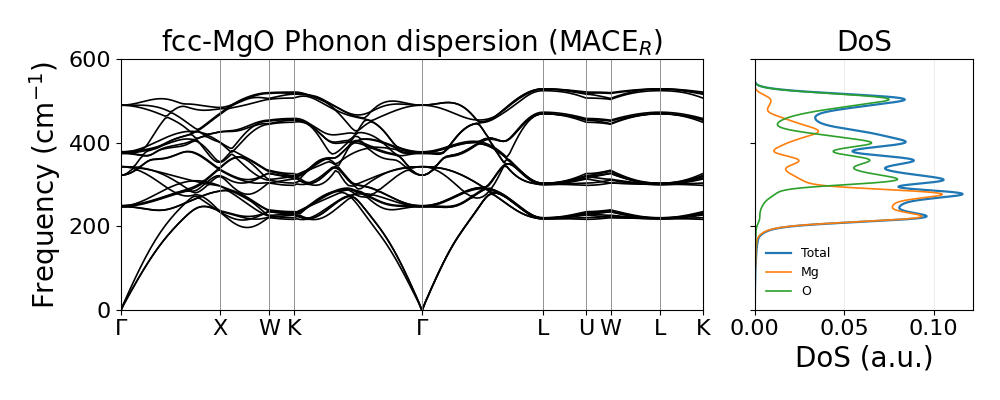}
        \put(1,35){\textbf{(c)}}
    \end{overpic}
    \par\vspace{1ex}
    \begin{overpic}[width=0.45\textwidth]{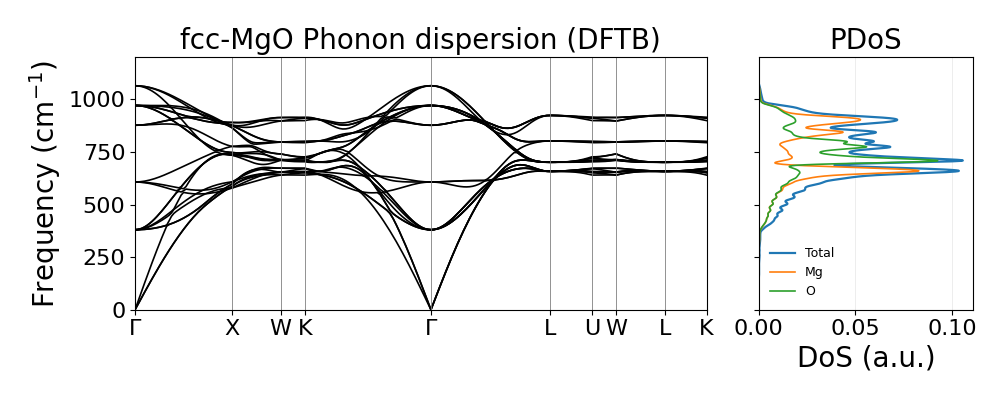}
        \put(1,35){\textbf{(d)}}
    \end{overpic}
    \caption{Phonon spectrum of fcc-\ch{MgO} calculated by (a) PBE-DFT, (b) DFTB+MACE, (c) $\text{MACE}_\text{R}$ and (d) DFTB. The moderate models are used for DFTB+MACE model and MACE$_{\text{R}}$. Note that the scale for DFTB is different from the scale of other two methods as the spectrum is much wider.}
    \label{fig:phononfccmgo}
\end{figure}

\begin{figure}[htbp]
    \centering
    \begin{overpic}[width=0.45\textwidth]{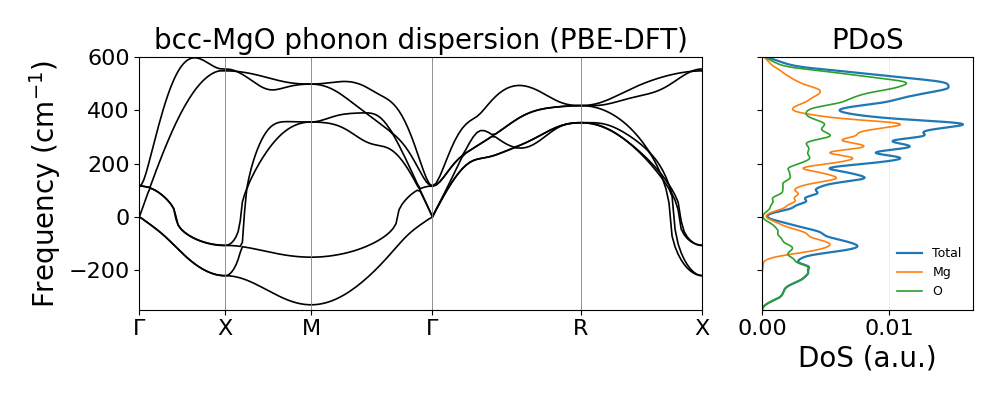}
        \put(1,35){\textbf{(a)}}
    \end{overpic}
    \par\vspace{1.6ex}
    \begin{overpic}[width=0.45\textwidth]{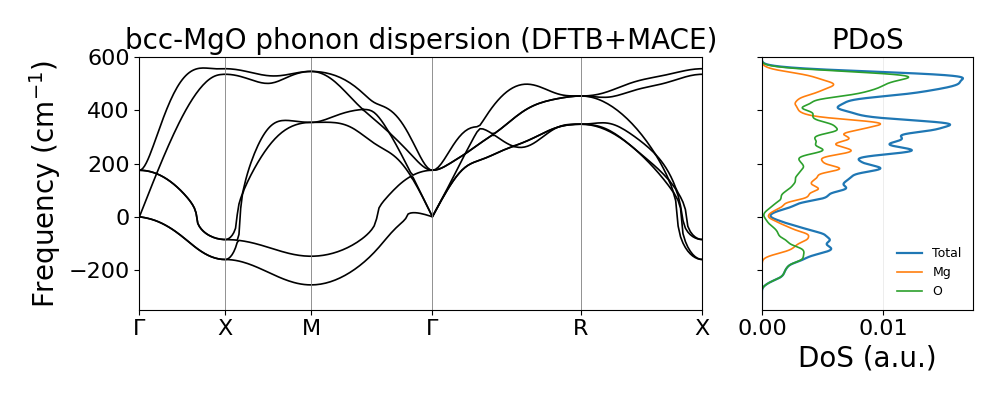}
        \put(1,35){\textbf{(b)}}
    \end{overpic}
    \par\vspace{1.6ex}
    \begin{overpic}[width=0.45\textwidth]{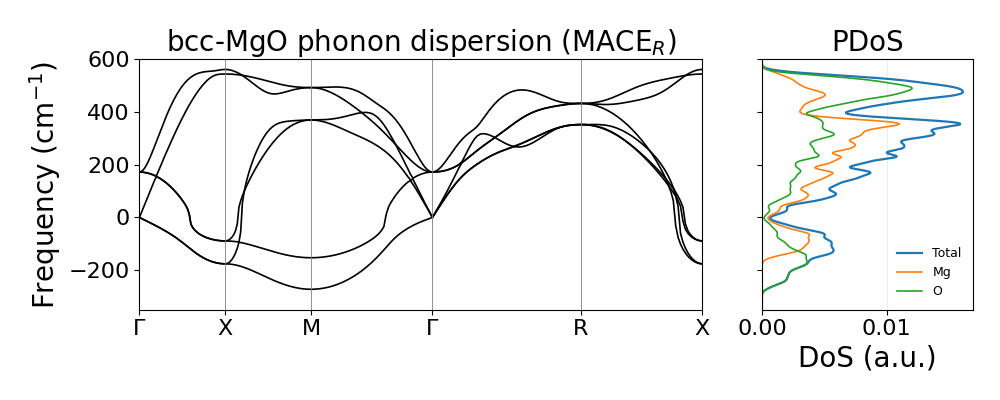}
        \put(1,35){\textbf{(c)}}
    \end{overpic}
    \par\vspace{1.6ex}
    \begin{overpic}[width=0.45\textwidth]{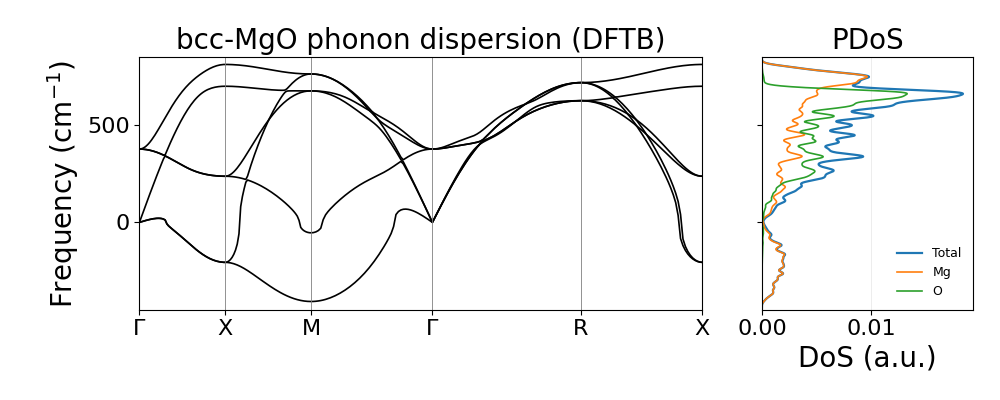}
        \put(1,35){\textbf{(d)}}
    \end{overpic}
    \caption{Phonon spectrum of bcc-\ch{MgO} calculated by (a) PBE-DFT, (b) DFTB+MACE, (c) $\text{MACE}_\text{R}$ and (d) DFTB. The moderate models are used for DFTB+MACE model and MACE$_{\text{R}}$. Note that the scale for DFTB is different from the scale of the other two methods.}
    \label{fig:phononbccmgo}
\end{figure}

Figure \ref{fig:phononfccmgo} shows the phonon spectra and Projected Density of States (PDoS) for phonons of fcc-\ch{MgO}, and figure \ref{fig:phononbccmgo} shows the phonon spectra and PDoS of bcc-\ch{MgO} calculated by PBE-DFT, DFTB+MACE, $\text{MACE}_\text{R}$ and DFTB. Introducing the MACE model to DFTB again leads to significant improvement in the phonon spectrum compared with DFTB only. For fcc-\ch{MgO}, the DFTB+MACE model performs better than $\text{MACE}_\text{R}$ for the acoustic phonons, as can be seen by comparing the band width of the phonon spectrum and the position where the phonon PDoS increases rapidly. This improvement behaves in the same way as for the bulk modulus. In the higher frequency region, DFTB+MACE still differs from the DFT reference and $\text{MACE}_\text{R}$, especially along the L-U-W-L-K path.

The limitations of DFTB alone are clearly shown throughout the phonon spectrum. The gradient of the acoustic branch is much higher than for PBE-DFT, indicating that there may be over-binding between \ch{Mg} and \ch{O}. The optical frequencies are also much higher than found by PBE-DFT, indicating that the bond stiffness is much higher than the PBE-DFT reference. 

For bcc-\ch{MgO}, the problems found for DFTB are the same. Even though it can correctly predict that the $\Gamma$ point is stable, the over-stiffness problem causes the optical branch frequency to be higher than the PBE-DFT reference. In addition, \ch{Mg} contributes to the imaginary part of the phonon PDoS. Although there is still a gap between the DFTB+MACE model and the PBE-DFT reference, it still shows the contributions of both \ch{Mg} and \ch{O} to the imaginary part and performs at a similar level as $\text{MACE}_\text{R}$. It should also be pointed that if we choose a small model, the DFTB+MACE model will still perform at a relatively acceptable level. In contrast, if the $\text{MACE}_\text{R}$ model loses the information from high irrep order, and the resulting predictions can be catastrophic. See figure 6 in the SI.

\begin{figure}[htbp]
    \centering
    \begin{overpic}[width=0.23\textwidth]{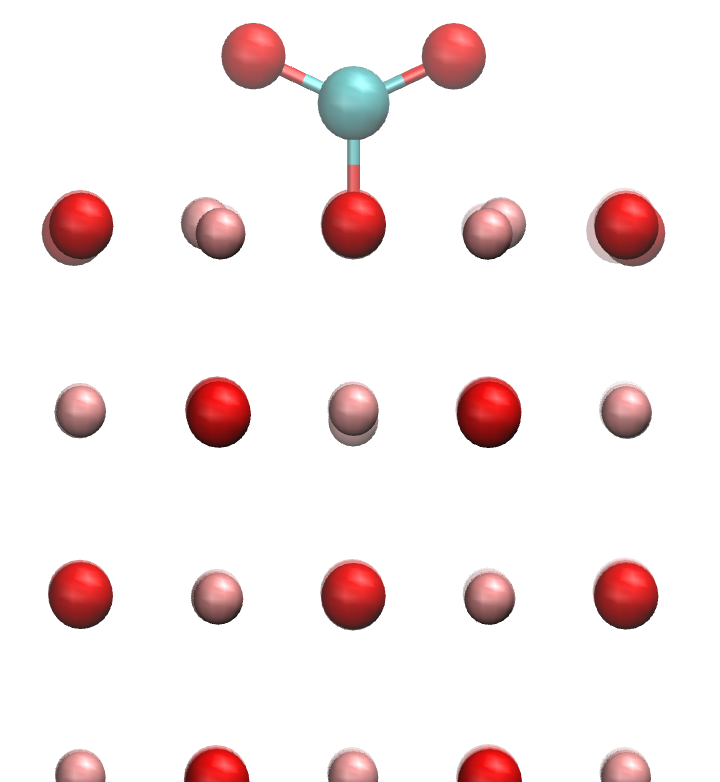}
        \put(1,85){\textbf{(a)}}
    \end{overpic}
    \hfill
    \begin{overpic}[width=0.23\textwidth]{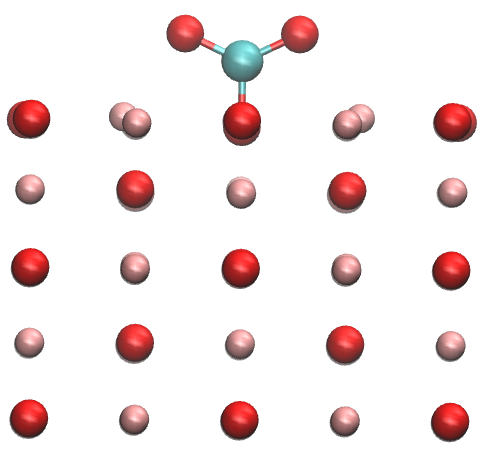}
        \put(1,92){\textbf{(b)}}
    \end{overpic}
    \caption{The structure for the adsorption of \ch{CO2} on the (a) \ch{MgO} (100) surface and (b) \ch{MgO} (110) surface calculated by the DFTB+MACE model. The coral balls represent \ch{Mg} atoms, the blue ball represents the \ch{C} atom and the red balls represent \ch{O} atoms.}
    \label{fig:mgoco2}
\end{figure}

Having shown the DFTB+MACE model performs well for bulk \ch{MgO}, we now move to surface and adsorption analysis. The (100) and (110) surfaces of \ch{MgO} and \ch{CO2} adsorption on these surfaces are now studied. In traditional DFTB it is challenging to describe the surface energy accurately. For example, when the surface is created, the electronic term is less stable or less negative than the bulk. However, the reduced coordination number caused by the surface also leads to a decrease in the number of pair potential contributions to the $E^{(0)}$ term. As a result, the parameterisation faces an inconsistency: it may be adjusted to reproduce bulk energies, but it may then incorrectly stabilize surface atoms. This reflects a fundamental limitation of environment-independent pair potentials in capturing both bulk and surface energies in the DFTB model. Table \ref{table2} shows, using the different models, the surface energies for the (100) and (110) surfaces of \ch{MgO}, the adsorption energy $E$ of \ch{CO2} on these two surfaces, the bond length between C and O in \ch{CO2} and in \ch{MgO}, and the bond angle of \ch{CO2}. Here a MACE foundation model (MACE-MP-0a) \cite{batatia2025foundation} result is also shown for comparison. Figure \ref{fig:mgoco2} shows the structure of \ch{CO2} adsorbed on the (100) and (110) surfaces of \ch{MgO} calculated using the DFTB+MACE model. It should be pointed that there is no periodic system with a surface in the training set: all surface information is from \ch{MgO} clusters, and the \ch{CO2}-\ch{MgO}(110) adsorption structure is not included in training set.

It can be seen that introducing MACE successfully resolves the apparent inconsistency and the surface energy is much more accurate compared with DFTB. The predicted value from the DFTB+MACE model for the (100) surface is larger than the PBE-DFT reference while for the (110) surface it is smaller. Both surface energies predicted by the $\text{MACE}_\text{R}$ model are smaller than PBE-DFT reference. The reason for the difference between DFTB+MACE and PBE-dFT is not that the non-periodic training set leads to an averaging effect in the learned model, because this does not happen in the $\text{MACE}_\text{R}$ model. The reason may be similar to what happened in bulk \ch{MgO}, where \ch{MgO} clusters do not cover all the chemical environments so the potential energy surface in the intermediate region is estimated. However, the result is still acceptable, and we note that the surface energy of the (110) surface happens to be closer to the experimental value 1.04 Jm$^{-2}$ \cite{jura1952experimental} than the DFT value, though this is accidental.

For the adsorption problem, all methods predict a relatively correct structure compared with the PBE-DFT reference. Unexpectedly, the $\text{MACE}_\text{R}$ model gave two positive predictions for adsorption energies, and the predictions of the DFTB+MACE model are about 1 eV more negative than the PBE-DFT reference. This clearly demonstrates the effect of estimating the potential energy surface in the intermediate region resulting from insufficient training data coverage, and the model needs to be fine-tuned for certain calculation. In addition, even though the standard MACE model provides high accuracy results for the \ch{MgO}'s (100) surface calculation, it is less accurate when predicting the \ch{MgO}'s (110) surface energy, and predicts an extremely repulsive adsorption energy and long \ch{C}-\ch{O} bond length, which implies there are not configurations with the required chemical environment even in the full training set. Although the DFTB+MACE model cannot completely eliminate this issue, it can significantly mitigate it, as the underlying electronic structure already provides a substantial amount of physical information. Given that the training set does not include different surface and adsorption configurations and $\text{MACE}_\text{R}$ model provides an opposite prediction, this result partly demonstrates the transferability of the DFTB+MACE model. However, higher accuracy will require more diverse surface and adsorption configurations be included in the training set. 

\begin{table*}[htbp]
    \caption{\label{table2} The surface energy $\gamma$($\text{Jm}^{\text{-2}})$, the adsorption energy $E_{\ch{CO2}-\ch{MgO}}$ (in eV), the bond length between \ch{C} in \ch{CO2} and O in \ch{CO2} and \ch{MgO} (Å), and the bond angle of \ch{CO2} (°) calculated by different models (see main text for details). Note that MACE-MP-0a is a foundation model \cite{batatia2025foundation}.}
    \begin{tabular}{c c c c c c}
         & PBE-DFT&MACE-MP-0a & DFTB+MACE& $\text{MACE}_{R}$& DFTB\\
        \hline
        $\gamma_{\ch{MgO}(100)}$&0.88&0.87&1.03&0.79&5.98 \\
        $E_{\ch{CO2}-\ch{MgO}(100)}$&-0.07&-0.39&-1.43&1.99&-5.62 \\
        $C-O(\ch{CO2})_{\ch{MgO}(100)}$&1.25&1.25&1.25&1.26&1.25 \\
        $C-O(\ch{MgO})_{\ch{MgO}(100)}$&1.49&1.48&1.45&1.45&1.32 \\
        $\angle OCO_{\ch{CO2}}^{\ch{MgO}(100)} $&135&136&132&136&128 \\
        \hline
        $\gamma_{\ch{MgO}(110)}$&2.14&7.05&1.80&1.64&6.37 \\
        $E_{\ch{CO2}-\ch{MgO}(110)}$&-2.59&-2.89&-3.34&2.19&-1.54 \\
        $C-O(\ch{CO2})_{\ch{MgO}(110)}$&1.28&1.21&1.27&1.25&1.25 \\
        $C-O(\ch{MgO})_{\ch{MgO}(110)}$&1.34&1.66&1.33&1.48&1.33 \\
        $\angle OCO_{\ch{CO2}}^{\ch{MgO}(110)} $&129&144&129&129&131 \\
    \end{tabular}
\end{table*}

\subsection{Molecular dynamics}

\begin{figure}[htbp]
    \centering
    \begin{overpic}[width=0.47\textwidth]{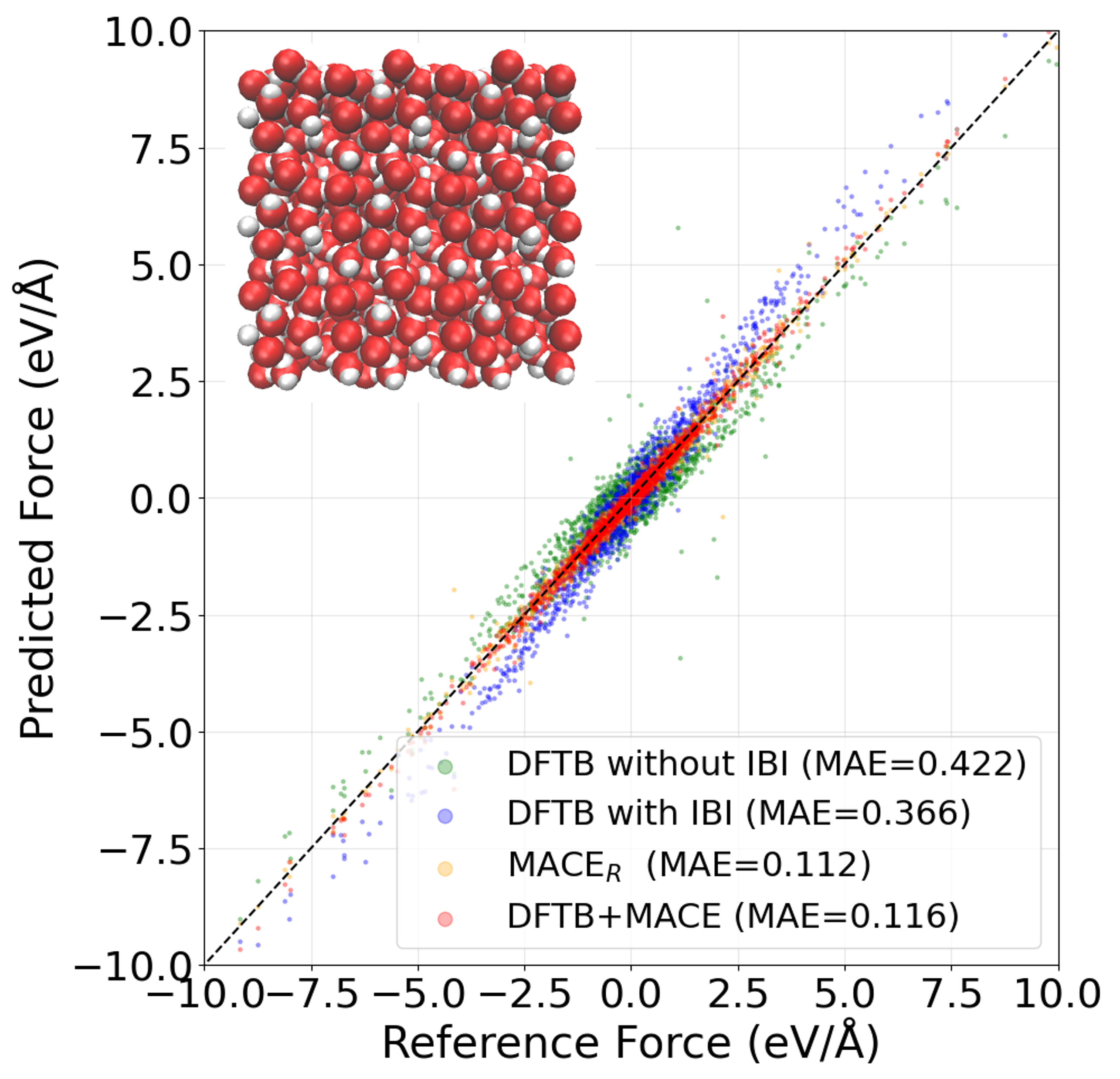}
        \put(0,95){\textbf{(a)}}
    \end{overpic}
    \par\vspace{1ex}
    \begin{overpic}[width=0.5\textwidth]{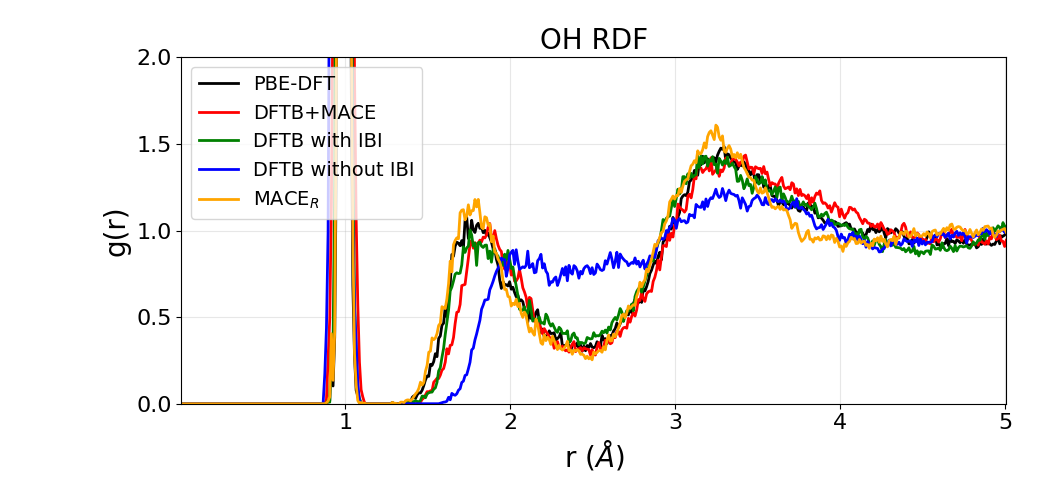}
        \put(2,35){\textbf{(b)}}
    \end{overpic}
    \par\vspace{1ex}
    \begin{overpic}[width=0.5\textwidth]{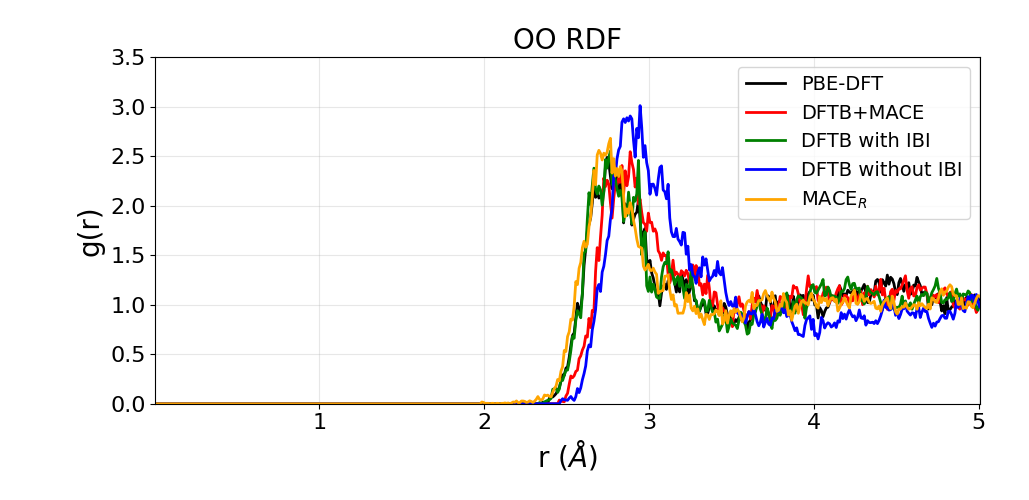}
        \put(2,35){\textbf{(c)}}
    \end{overpic}
    \caption{(a) The initial geometry of a water bulk and the predicted force value against the reference value for different models (note that the scale is different from the scale in figure \ref{fig:R}).  The radial distribution function (RDF) of (b) \ch{OH} and (c) \ch{OO} produced by different models. The moderate models are used for DFTB+MACE model and MACE$_{\text{R}}$.}
    \label{fig:rdf}
\end{figure}

We now carry out molecular dynamics simulations of bulk water and water on the (0001) surface of \ch{Mg(OH)2} to investigate finite temperature effects. It is hard to reproduce the correct structure of bulk water with a short ranged pair potential, therefore in traditional DFTB an iterative Boltzmann inversion (IBI) method (also known as reverse Monte Carlo) is usually adopted. \cite{lyubartsev1995calculation, doemer2013situ,goyal2014molecular,lourenco2020accurate}  It introduces a correction in the long range region so that the pair potential exhibits alternating attractive and repulsive oscillations, rather than a monotonically decaying behaviour. Figure \ref{fig:rdf} (a) shows the initial geometry of water bulk and the atomic forces from different models compared with PBE-DFT reference forces. Figure \ref{fig:rdf} (b) and (c) show the radial distribution function (RDF) of \ch{OH} and \ch{OO} generated by different methods. The DFTB+MACE model reaches an accuracy of 116.0 meV/Å, which is similar to that of $\text{MACE}_\text{R}$. The force accuracy of DFTB with IBI corrections is higher than DFTB without IBI corrections, but the MAE still remains at a high level. It can be found that all methods, except DFTB without any corrections, show good agreement with the PBE-DFT reference RDF: it can clearly describe the correct position of the first \ch{OH} peak around 1.8 Å, the second \ch{OH} peak around 3.2 Å and the first \ch{OO} peak around 2.8 Å. In our previous study\cite{yu2025tight}, it has been proven that after around 4 to 5 iterations of the IBI method, the RDF curve calculated by DFTB with the IBI correction can be almost exactly the same as the reference data. However, even though the IBI method can provide an even better RDF curve than the DFTB+MACE model, different from what had been expected, there is no dramatic improvement in the forces of DFTB with the IBI correction (from 421.5 meV/Å to 366.2 meV/Å) compared with DFTB without the IBI correction. The RDF can provide structural information but it cannot determine the forces which are related more to atomic vibrations and diffusion. With a simple $\text{MACE}_\text{R}$ model it is hard to provide the correct second \ch{OO} peak. This may reflect the intrinsic limitation of a simple MACE model, as it has been reported that MACE, or other MLIPs, are struggling to accurately describe long-range interactions \cite{king2025machine,gao2025foundation}. Ideally, the long-range interaction generated from DFTB can compensate for this error, which requires a more precise parameterisation for the electronic part.

\begin{figure}[htbp]
    \centering
    \begin{overpic}[width=0.47\textwidth]{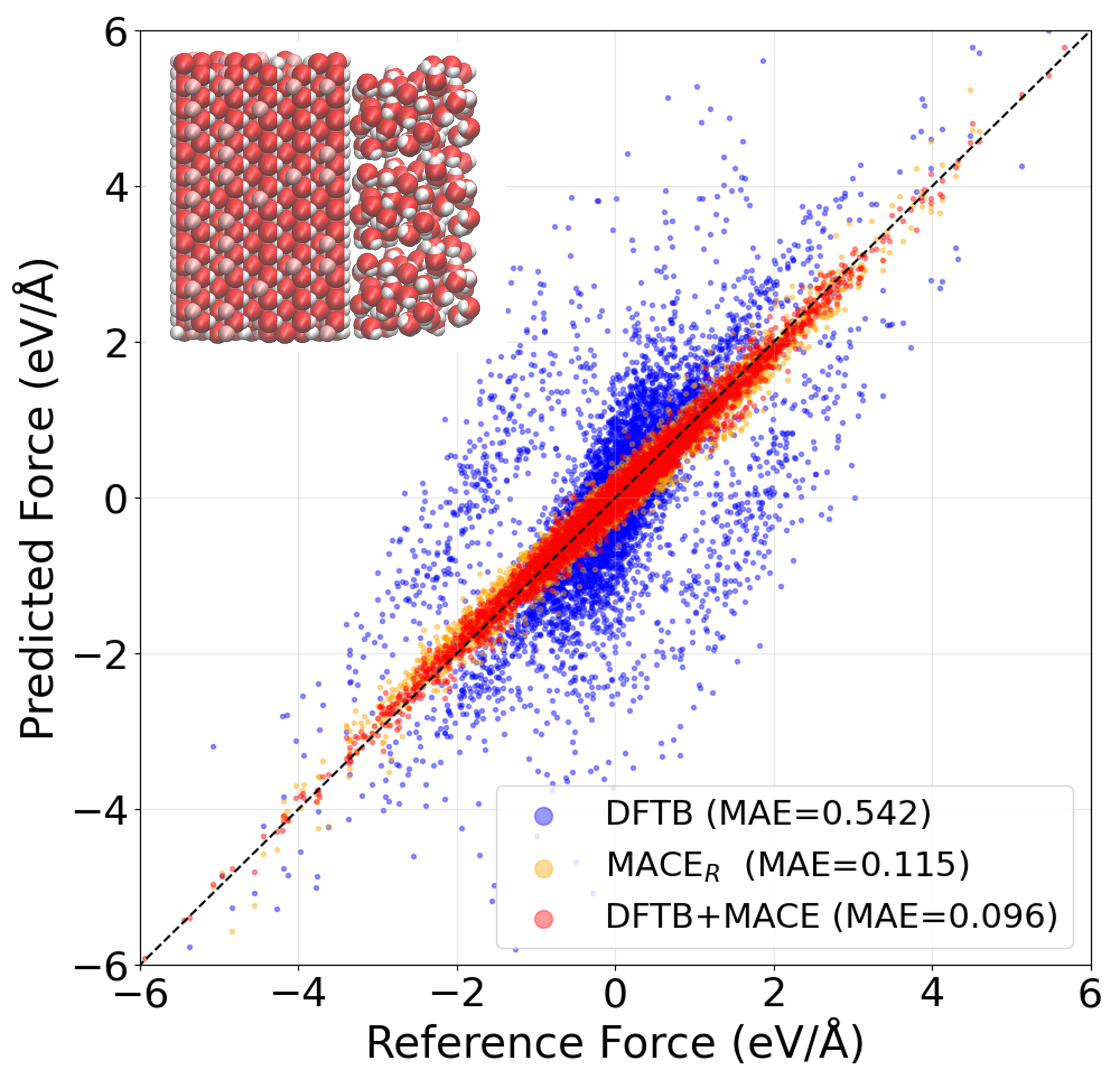}
    \end{overpic}
    \caption{The initial geometry of a water bulk on (0001) surface of \ch{Mg(OH)2} and the predicted force value against the reference value for different models. The moderate models are used for DFTB+MACE model and MACE$_{\text{R}}$. Note that the scale is different from the scale in figure \ref{fig:R}.}
    \label{fig:mgohh2o}
\end{figure}

The computational cost for \textit{ab initio} Molecular Dynamics (MD) of water bulk on the (0001) surface of \ch{Mg(OH)2} is extremely high because of the large number of atoms: for example, around 4 hours are required for one step using a 64 core Intel Xeon Platinum 8358. Thus, we randomly selected five snapshots, recalculate the energy and forces using DFT and compare the results of DFTB, MACE$_\text{R}$, DFTB+MACE with the PBE-DFT reference. Figure \ref{fig:mgohh2o} shows the initial structure of a water bulk on a (0001) surface of \ch{Mg(OH)2} and the comparison between the DFTB+MACE results and PBE-DFT reference. The MAE of forces is only 96.2 meV/Å, which is even surprisingly better than the validation MAE. This is a dramatic improvement compared with traditional DFTB, which has a force MAE of 542.1 meV/Å. For bulk water on a brucite surface, the DFTB+MACE model is much better than DFTB: the IBI method may have adverse effects here because the chemical environment around the surface is extremely different from that in pure bulk water.

It should also be noted that, for the DFTB+MACE model, the structure may be trapped in a local minimum that depends on the initial structure, and the final structure can depend on the initial configuration. A possible reason is that the current training set was originally constructed for a standard MACE model rather than specifically for the DFTB+MACE framework. The training data are concentrated around equilibrium structures because the global minimum of the potential energy surface corresponds to the physical equilibrium configuration. In contrast, the DFTB+MACE model learns a correction to the DFTB repulsive term. The minimum of the repulsive potential does not coincide with the equilibrium structure because the electronic contribution varies monotonically with bond length or lattice constant. In our previous work, the minima of the pairwise repulsive potentials were found at 1.28 Å for O–O, 0.77 Å for H–H, and 1.06 Å for O–H interactions. However, configurations near these distances occur only rarely in molecular or crystalline equilibrium structures and therefore are poorly represented in the current training set (see Supporting Information figure 2-4). As a result, the model receives limited information in regions that are particularly important for describing the repulsive potential. The system may become trapped in local minima that do not appear in the PBE-DFT reference during geometry optimization. Future work should therefore focus on constructing training datasets specifically for the DFTB+MACE framework. Such datasets should not be restricted to equilibrium configurations but should provide broader coverage of the potential energy surface, particularly in the vicinity of the repulsive-potential minima and other short-range interaction regions.

\section{Conclusion and Future Work}
In conclusion, we integrate the MACE potential into the DFTB framework by replacing the traditional pairwise repulsive term with a MACE potential. This model balances the computational costs and accuracy, and is able to provide electronic structure. The presence of the electronic term ($E^{(1)}$ and $E^{(2)}$), together with the isotropic atomic reference density $n^{(0)}_\alpha$ assumption for $E^{(0)}$ , enables the DFTB+MACE model to achieve high accuracy with reduced MACE model complexity compared to the $\text{MACE}_\text{R}$ model alone. Only in more complex models does $\text{MACE}_\text{R}$ model slightly outperform DFTB+MACE model in accuracy. The possible reasons include that the model has reached its limitation, the errors in DFTB electronic terms are difficult for MACE to reproduce, and the training set does not cover enough potential surface. There is no training set currently designed for a DFTB+MACE model, while the quality of model depends strongly on training set \cite{wu2026hitchhikers}. Improving the quality of DFTB parameters related to the electronic structure for the magnesium study will be part of our future work.

The trained models were applied to structural relaxation, phonon calculations and molecular dynamics calculations. The DFTB+MACE model consistently showed improved accuracy with a moderate hyperparameter setting in the MACE potential compared with existing DFTB parameter sets. For structures with complex chemical environments, such as adsorption problems, DFTB+MACE delivers better trends, transferability and numerical results compared with traditional DFTB, but there still remains a gap in numerical accuracy compared with the DFT reference. During the molecular dynamics calculations, there is a large improvement for the DFTB+MACE model compared with traditional DFTB. Not only the structure, but also the force, which is more important for other studies such as atomic vibration and diffusion, show encouraging results. 

Future work will focus on two main directions. A dedicated training dataset should be developed for the DFTB+MACE framework, with broader coverage of the potential surface. Second, the quality of the underlying electronic structure description will be further improved to provide more accurate energies and forces. Overall, for solid-state systems, the DFTB+MACE approach significantly improves the accuracy of energies and forces while introducing only a modest computational overhead. At the same time, it retains the electronic information available in the DFTB framework, making it a promising method for large-scale atomistic simulations.

\section*{Supporting Information}
The supporting Information is available at \url{to_be_filled}.

\section*{Data Availability Statement}
The DFTB+MACE PLATO patches and the training set for both DFTB+MACE and MACE model are available via \url{https://github.com/d3iven005/PLATO-tools}. PLATO package can be made available on request to the authors.
\begin{acknowledgments}
This research was funded by the Department of Materials, Imperial College London. The DFT calculations of training set and machine learning calculations were supported by the Imperial College Research Computing Service, DOI: 10.14469/hpc/2232. We acknowledge the Thomas Young Centre under grant number TYC-101.
\end{acknowledgments}

\section*{Author Contributions}
\textbf{Jiwen Yu}: Conceptualization (equal); Writing-original draft (equal); Software (equal); Data analysis (equal).
\textbf{Arash A Mostofi}: Conceptualization (equal); Writing-review \& editing (equal); Data analysis (equal).
\textbf{Andrew Horsfield}: Conceptualization (equal); Software (equal); Writing-review \& editing (equal); Data analysis (equal).

\nocite{*}
\bibliography{aipsampref}

\end{document}